\documentclass[11pt]{article}
\usepackage[nosort]{cite}
\input cyracc.def

\usepackage{amsmath,amsthm,amsfonts,amssymb,amscd,mathrsfs,slashed,graphicx,braket}

\newlength{\xtrawidth}
\newlength{\xtraheight}
\numberwithin{equation}{section}
\numberwithin{table}{section}
\numberwithin{figure}{section}
\newcommand{\dd}{\mathrm{d}}

\newcommand{\num}{{}}

\usepackage{xcolor}
\usepackage{arydshln}
\usepackage[font=footnotesize]{caption}
\usepackage{float}
\usepackage{braket}
\usepackage{extarrows}
\usepackage{bbm}
\usepackage{nicefrac}
 \usepackage{import}
 \usepackage{bbold}
\usepackage{mathrsfs}
\usepackage{pgf,tikz}
\usetikzlibrary{arrows}
\usepackage[fleqn]{mathtools}
\usepackage{graphicx}
\usepackage{subcaption}
\usepackage{dsfont}
\usepackage{enumerate}
\usepackage{enumitem}
\usepackage{nccmath}
\usepackage{booktabs}

\newenvironment{gleichung}{\begin{equation}\begin{aligned}}{\end{aligned}\end{equation}\\ \noindent}
\newenvironment{gleichung*}{\begin{equation*}\begin{aligned}}{\end{aligned}\end{equation*}}

\renewcommand{\bf}{\textbf}

\newcommand{\url}{{}}

\begin{document}

\begin{titlepage}

\begin{center}
\hfill BONN--TH--2019--07\\
\vskip 0.6in
{\LARGE\bf{The $l$-loop Banana Amplitude from  GKZ Systems \\[0.9ex] and relative Calabi-Yau Periods}}

\vspace{20mm}
{
Albrecht Klemm\footnote{aklemm@th.physik.uni-bonn.de}, 
Christoph Nega\footnote{cnega@th.physik.uni-bonn.de },
Reza Safari\footnote{rsafari@th.physik.uni-bonn.de}\\[4mm]
{\textit{\textsuperscript{123}Bethe Center for Theoretical Physics and \textsuperscript{1}Hausdorff Center for Mathematics,\\
{Universit\"{a}t Bonn, D-53115 Bonn}}} }\\[2mm]

\end{center}

\vspace{2cm}
\begin{center}
\textbf{Abstract}
\end{center}
We use the GKZ description of  periods and certain classes of relative periods  on families of 
Barth-Nieto Calabi-Yau $(l-1)$-folds in order to solve the $l$-loop banana amplitudes with their 
general mass dependence. As  examples we compute the  mass dependencies of the banana amplitudes 
up to the three-loop case and  check the results against the known results for special mass values.

\end{titlepage}

\section{Introduction}

It was observed in~\cite{MR1011353} by Gel\cprime fand, Kapranov and Zelevinsk\u{\i} (GKZ) that
 {\sl practically all  integrals that arise in perturbative quantum field theory} have the form of residuum 
 integrals of rational functions defined in a toric variety $\mathbb{P}_\Delta$. We will call these the  GKZ 
period integrals. In applications the more relevant statement  is that in dimensional regularization say
in  $4-2 \epsilon$ dimensions the coefficients of the Laurent expansion of the Feynman integral  in $\epsilon$ are such period 
integrals~\cite{Bogner:2007mn}. 

The simplest GKZ integrals are related to the Griffiths residuum form~\cite{MR229641,MR233825} of 
geometric  forms in the cohomology of varieties $M$ that are algebraically embedded in the toric varieties.  
The period integrals over closed cycles are solutions to a system of linear homogeneous differential 
operators called the Picard-Fuchs differential ideal (PFDI). The main result of GKZ is that the 
GKZ integrals are determined by a system of linear differential operators, w.r.t. to their parameters,  
defined  in~\cite{MR1011353,MR1020882,MR1080980} and called the GKZ system.  This is of 
course only true up to linear combinations of the solutions, which reflects the choice of the  
homology class of the integration domain.  This  GKZ system can be thought  as a generalization 
of the hypergeometric systems. 
  
It is related in simple cases in which $M$ is a Calabi-Yau manifold to the PFDI as follows.  After 
pulling out a moduli dependent multiplicative factor --- the coefficient of the unique inner point in the 
Newton polyhedron ---  from the GKZ integrals, the {\sl geometric period integrals}  that solve the PFDI are among the solutions of the 
modified GKZ system~\footnote{As explained in \cite{Hosono:1993qy,Hosono:1995bm} the PFDI 
can be obtained from the modified GKZ system by factoring it from the latter.}\cite{Hosono:1993qy,Hosono:1995bm}. 
The application of~\cite{Hosono:1993qy,Hosono:1995bm} shows that for many 
problems the PFDI  is much easier  obtained from the GKZ then from the Griffiths 
reduction method. The latter is a generic algorithm that produces differential 
relations between the periods by  chains of partial integrations, while the former 
uses simple symmetries of the integrand reflecting symmetries in the 
parameter space.  More generally, the GKZ integrals can involve 
non necessary holomorphic forms that are integrated  over chains in $M$~\cite{MR0260733}. 
The latter are called relative periods and fulfill an inhomogeneous system of 
differential operators. It turns out that the  solutions of  GKZ systems can be related to 
periods as well as to a class of relative periods.    

In this note we consider particular Feynman integrals that correspond to a class of $l$-loop 
Feynman diagrams in two space-time dimensions with two vertices of valence $l+1$, one invariant momentum $K^2$, 
and  $l+1$  different masses $M_i$ for each propagator, known as Banana diagrams. These are  
depicted in Figure \ref{bananadiagram}. By the dimensional shift relations these integrals yield the leading terms in the dimensional  
regularization parameter $\epsilon$ in $4-2 \epsilon$ dimensions\cite{Tarasov:1996br,Lee:2009dh}.  
For these integrals\footnote{Also for other Feynman graphs the appearing integrals can be related to Calabi Yau integrals as pointed out in~\cite{Bourjaily:2018ycu,Bourjaily:2018yfy,Bourjaily:2019hmc}.} given in  (\ref{bananageneral}) the numerator of the rational function is trivial and the 
homogeneous differential system 
is related to the PFDI for the closed periods of the holomorphic $(l-1,0)$-form 
of a Calabi-Yau $(l-1)$-fold $M$.  However, the integration domain of the Feynman 
integral has  in general boundaries. It turns out that  the solutions for the closed 
periods describe only the maximal cut integral of the Feynman amplitude\footnote{See also \cite{Primo:2016ebd} for a connection between maximal cut Feynman integrals and solutions to corresponding differential equaions.}. The latter is  an 
important building block for the  description of the physical amplitude, which however
has to be supplemented by special solutions to the inhomogeneous equations, which 
correspond to the boundary contribution of the relative periods. 

A fact of great importance for mirror symmetry is  that Calabi-Yau manifolds are expected to have 
at least one point of maximal unipotent monodromy in their moduli space. For Calabi-Yau
hypersurfaces and complete intersections in toric ambient spaces, the  location of 
these points can in fact be calculated  purely combinatorial from triangulations of the  toric 
polyhedron.  At such a  point the local exponents for the solutions of the PFDI are completely 
degenerate. A consequence is  that there is an unique analytic solution, while all 
other solutions are all logarithmic at this point. There is also an unique solution with the highest power of 
logarithms  which equals the dimension $l-1$ of $M$. Moreover, the maximal cut integral corresponds 
to the unique holomorphic period and can be evaluated directly by a residuum integral over an $l$-dimensional 
torus  in  $\mathbb{P}_\Delta$.  All logarithmic closed periods can be obtained by the Frobenius method.  
One of the technical insides of this paper is that also the inhomogeneous solutions can be constructed at this point
from symmetries of the  GKZ system and the explicit form of the GKZ integral and we show that this 
method\footnote{See also~\cite{delaCruz:2019skx,Klausen:2019hrg} for a different discussion of the GKZ system in the context of Feynman integrals with generic mass dependencies.}  is practical  enough to calculate the  full mass  dependence for the three-loop amplitude and 
maybe beyond.            

Only for the two-loop amplitude the dependence on all  three parameters  has been calculated so 
far\footnote{In the two-loop case the diagram is also called sunset diagram.} in~\cite{MR3780269}. The knowledge of the general mass dependence is not only important from a conceptual point but it is also required in the computation of higher loop corrections to certain processes studied at the Large Hadron Collider, as for example in Higgs production processes~\cite{Bonciani:2019jyb}.
It turned out that the integral is closely related to the period integral of the local mirror $M$ of the 
non-compact Calabi-Yau three-fold $W$ defined  as total space of the anti-canonical line bundle over 
the degree three del Pezzo surface $S$, which is $\mathbb{P}^2$ blown 
up in three generic points. The masses are related in a simple way  to the three new K\"ahler 
parameters in the blown up geometry and the toric polyhedron representing  $S$ in 
Figure \ref{toricl=2}.  By  local mirror symmetry the toric polyhedron is also the Newton 
polyhedron for the polynomial in the denominator of the GKZ integral, whose vanishing 
locus  is a special family of elliptic curves ${\cal E}$, i.e. the Calabi-Yau one-fold.   
The period problem of the meromorphic differential of the third kind, whose non-vanishing residua 
correspond to the masses, on this elliptic curve has been solved universally for all 
toric del Pezzo surfaces in terms of modular forms~\cite{maximilianalbrecht}. It contains the 
information of the maximal cut integral. 

The elliptic curve above will be replaced by a $\mathrm K3$ for the three-loop case and a Calabi-Yau $(l-1)$-fold 
for the $l$-loop case. This banana diagram with general mass dependencies plays an important role in three-loop corrections to the $\rho$ parameter where the top and bottom quark masses are considered~\cite{Abreu:2019fgk}. For the  four-loop case the Calabi-Yau three-fold takes the form of a mass deformation 
of the  one parameter family of Barth-Nieto quintics. Their form can be readily generalized to arbitrary dimensions 
as in equation (\ref{BarthNieto}) and will be called Barth-Nieto Calabi-Yau $(l-1)$-folds and describe the geometry
of the  $l$-loop graph (\ref{bananageneral}) for equal masses\footnote{Due to an additional scaling freedom we can actually set in the equal mass case all masses to unity.} $\xi_i=M_i/\mu=1$ for all $i$, depending only on 
$t=K^2/\mu^2$. For the  Barth-Nieto Calabi-Yau $(l-1)$-folds the $(l+1)$'th order Picard-Fuchs differential operator ${\cal D}^{(l+1)}_t$ 
is easily obtained. One  evaluates the geometric integral  (\ref{bananageneral}) with $\xi_i=1$ over  the 
$(l-1)$-torus $T^{l-1}$ instead of $\sigma_l$. The latter is readily performed  as it is equivalent to  an 
integral  over an $T^{l}$  torus in the ambient space, which leads by a simple residue 
calculation to an explicit  power series $\varpi(t)$.  ${\cal D}^{(l+1)}_t$ can only have rational coefficients and 
demanding that it annihilates $\varpi(t)$  fixes it uniquely~\cite{verrill1996}.
More efficiently is the method proposed in~\cite{Vanhove:2014wqa} using a decomposition of the integral in terms of  Bessel functions~\footnote{They are given up to $l=5$ together with a computer program that calculates 
them quickly for higher $l$ ~\cite{Vanhove:2014wqa}.}.

The full set of solutions to the Picard-Fuchs differential ideal and many aspects of their monodromies and 
analytic continuations have been intensively studied using the GKZ system in the context of mirror symmetry 
for period integrals of the  holomorphic $(n,0)$-form for  Calabi-Yau n-folds. For compact Calabi-Yau three-folds 
realized as hypersurfaces this was done in~\cite{MR1269718,Hosono:1993qy,Hosono:1995bm} or  as complete 
intersections~\cite{Hosono:1994ax,MR1328251,MR1463173} in toric varieties. Higher dimensional 
Calabi-Yau spaces  have been studied in \cite{Greene:1993vm,Mayr:1996sh,Klemm:1996ts,Bizet:2014uua,Gerhardus:2016iot}. An introduction 
and overview can be found in~\cite{MR3965409}. 
 
Our paper is structured as follows: In section \ref{subsec:lloop} we introduce the $l$-loop banana graphs and explain their geometric interpretation. To these Feynman diagrams we associate $(l-1)$-dimensional Calabi-Yau hypersurfaces. Their definition and useful properties are discussed in section \ref{ssec:geometry} and \ref{ssec:CalabiYauintoric}. In sections \ref{periodsgkz} to \ref{ssec:geomperiods} we introduce the notion of periods and relative periods on Calabi-Yau hypersurfaces and explain the GKZ method. Moreover, we explain the restriction on the physical subslice. The extension of the GKZ system to relative periods is developed in \ref{inhomstrategy}. This is the main part of our approach of computing $l$-loop banana amplitudes. In section \ref{examples} we calculate with our approach three examples, namely the one-, two- and three-loop banana graph. Finally, we make our conclusions in section \ref{conclusions}.

\vskip 1cm

\noindent
{\bf {Note added in draft:}} While we were in the process of finishing this draft an interesting paper \cite{Feng:2019bdx} has been published about evaluating Feynman integrals and GKZ method which seems to have some overlap with our work.

\section{$l$-loop Banana Diagram in the Toric Approach}
\label{chapgeneral}
We give the $l$-loop banana diagrams a geometric interpretation enabling us to use toric geometry to evaluate them. This geometric interpretation originates from the graph polynomial representation of a Feynman diagram which is obtained after Feynman parametrization and evaluation of many Gaussian integrals\footnote{For a review of the graph theoretical representation of Feynman diagrams we refer to \cite{Bogner:2010kv}.}. For the banana type diagrams in two dimensions the exponent of the first Symanzik polynomial vanishes and the exponent of the second Symanzik polynomial is one. This simplifies the form of these integrals a lot and offers us a geometric interpretation inspired from string theory. We regard the denominator of the integrand as a Newton polynomial which defines a Calabi-Yau hypersurface. The corresponding  banana diagram is viewed as a relative period of this Calabi-Yau hypersurface. Through the GKZ system of differential equations we construct a basis of periods on the Calabi-Yau variety at the maximal unipotent monodromy point. Extending the GKZ system to inhomogeneous differential operators we can write down a complete set of functions parametrizing the full banana amplitude.

\subsection{$l$-loop banana diagram}
\label{subsec:lloop}

\begin{figure}[h!]
	\centering
	\includegraphics[width=0.6\textwidth]{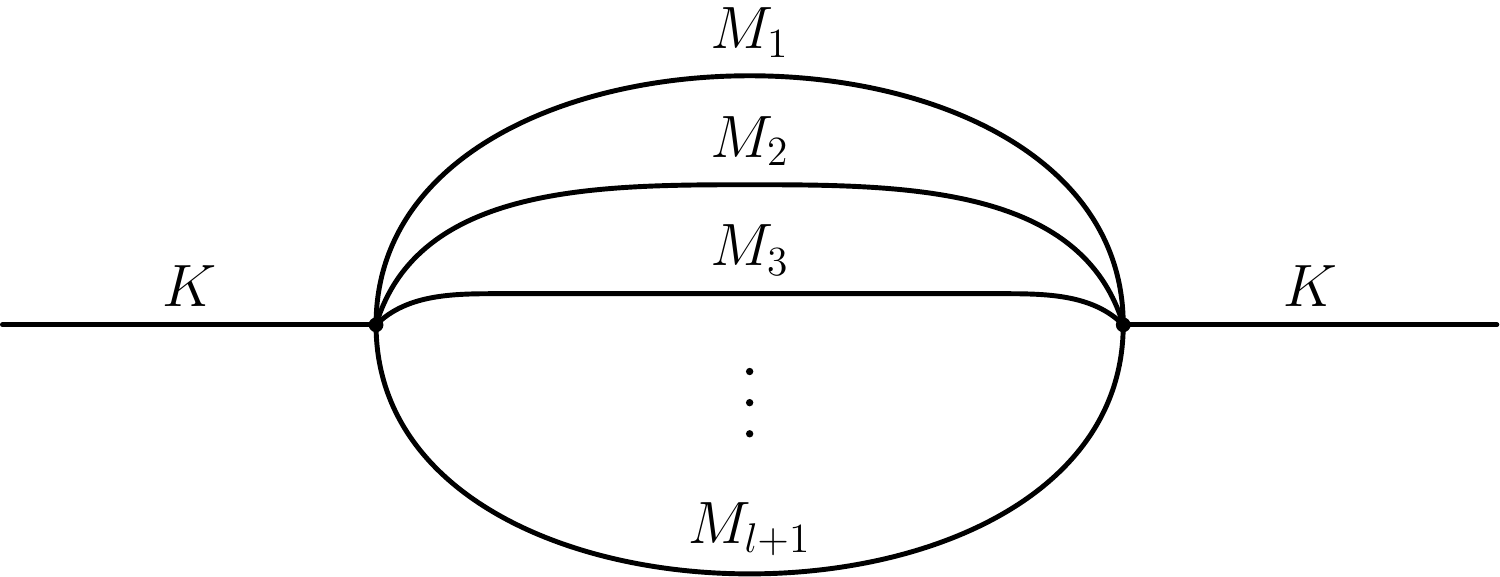}
     \caption{The l-loop banana diagram}
     \label{bananadiagram}
\end{figure}

The Feynman integral related to a $l$-loop banana diagram drawn in Figure \ref{bananadiagram} of a  2d  QFT with the 
corresponding interactions is given in  Feynman parametrization as
\begin{gleichung}
{\cal F}_{\sigma_l}(t,\xi_{i})=\int_{\sigma _{l}} \dfrac{\mu_{l}}{P_l(t,\xi_i;x)}=\int_{\sigma _{l}} \dfrac{\mu_{l}}{\left(t-\left(\sum_{i=1}^{l+1} \xi_i^2 x_i\right) \left(\sum_{i=1}^{l+1} x_i^{-1}\right) \right)\prod_{i=1}^{l+1} x_i  }~.
 \label{bananageneral}
\end{gleichung}
Here $x_i$ are the homogeneous coordinates of $\mathbb{P}^l$ and the $l$ real dimensional 
integration domain $\sigma_l$ is defined as 
 \begin{gleichung}
 \sigma_l=\{(x_1:\ldots: x_{l+1}) \in \mathbb{P}^l | x_i\in \mathbb{R} \ \text{with} \ x_i \ge 0, \ \forall i\}  \ .  
\label{sigmal}
\end{gleichung}  
The holomorphic $l$ measure $\mu_l$ is 
\begin{gleichung}
\mu_{l} =\sum_{k=1}^{l+1} (-1)^ {k+1} x_k \dd x_1 \wedge  \ldots \wedge {\widehat {\dd x_k} } \wedge  \ldots \wedge \dd x_{l+1}~ .
\label{measure}
\end{gleichung}
The parameters or moduli in (\ref{bananageneral}), $t$ and $\xi$, are dimensionless:  $t=\frac{K^2}{\mu^2}$ and $\xi=\frac{M_i}{\mu}$ for $i=1,\ldots l+1$, 
where $K$ is the external momentum, $M_i$ are the $l+1$ masses and $\mu$ is an infrared scale.                              

The key observation discussed more in the next subsection  is that (\ref{bananageneral}) is the 
GKZ period integral for a Calabi-Yau hypersurface in a toric ambient space.

\subsection{Geometry associated to $l$-loop banana diagram}
\label{ssec:geometry} 

The zero locus of the denominator of the integral defines a singular family of $(l-1)$-fold Calabi-Yau hypersurfaces $M_s$ as
\begin{equation}
M_s=\left\{ P_l(t,\xi_i;x)=0 | (x_1: \ldots : x_{l+1}) \in \mathbb{P}^l\right\} \ . 
 \end{equation}     
Due to standard arguments, see e.g. \cite{MR658304}, $M_s$  is a complex K\"ahler manifold  with trivial 
ca\-no\-ni\-cal class $K=0$,  hence a Calabi-Yau space. The first fact  follows by the definition of $M_s$ as 
hypersurface in projective space  $\mathbb{P}^l$ and the second as  for a  homogeneous polynomial  $P_l$
of degree  deg$(P)$  in $\mathbb{P}^l$  the canonical class is given in terms of the hyperplane class $H$  
of  $\mathbb{P}^l$ as~\cite{MR658304}  $-K=c_1(TM_s)=[(l+1)- {\rm deg}(P)]H$ and deg$(P)=(l+1)$. Note 
that given the scaling  of (\ref{measure}) this degree   makes the integrand of (\ref{bananageneral}) well 
defined under the $\mathbb{C}^*$ scaling  of the homogenous coordinates  defining  $\mathbb{P}^l$. 
Embedded in  $\mathbb{P}^l$ the hypersurface is a singular Calabi-Yau space.  
Due to the Batyrev construction there is a canonical resolution of these singularities to define a 
smooth Calabi-Yau family,  which  we discuss next following~\cite{MR1269718,Hosono:1993qy,MR1328251,MR3965409}.   
A Calabi-Yau  manifold $M$ of complex dimension $n=l-1$ has two characteristic global differential forms. 
Since it is K\"ahler it has a K\"ahler $(1,1)$-form $\omega$  defining its K\"ahler---  or symplectic structure deformations  space. 
The triviality of the canonical class implies the existence of  an unique holomorphic  $(n,0)$-form that plays a crucial role in the 
description of the  complex structure deformations space of $M$.

\subsubsection{Calabi-Yau hypersurfaces in toric ambient spaces}
\label{ssec:CalabiYauintoric}

First we define a Newton polynomial $P_{\Delta_l}$  as
\begin{align}
P_l(t,\xi_i;x)= :P_{\Delta_l}  \prod_{i}^{l+1}x_{i} ~.
\label{Newtonpolyhedron}
\end{align}
The exponents of each monomial of $P_{\Delta_l}$, w.r.t. to the  coordinates $x_i$, $i=1,\ldots,l+1$,   
define  a point in a lattice $\mathbb{Z}^{l+1}$. The convex hull of all these points in the natural 
embedding  of  $\mathbb{Z}^{l+1}  \subset \mathbb{R}^{l+1}$ defines an $l$-dimensional lattice 
polyhedron. The dimension is reduced due to the homogeneity of $P_{\Delta_l}$ and we denote
the polyhedron\footnote{One calls $P_{\Delta_l}$ the Newton polynomial of $\Delta_l$ and $\Delta_l$ the Newton 
polyhedron of $P_{\Delta_l}$.}   that lies in the induced lattice $\mathbb{Z}^{l}\subset \mathbb{R}^{l}$ by $\Delta_l$.

More concretely, picking the canonical  basis $e_i$  for $\Lambda= \mathbb{Z}^{l}\subset \mathbb{R}^{l}=\Lambda_{\mathbb{R}}$  
the  $l(l+1)$ vertices  defined by (\ref{bananageneral}) and (\ref{Newtonpolyhedron})  span the polytope $\Delta_l$~\footnote{For $l=1,2,3$ 
these polytopes are depicted in Figures \ref{toricl=1},   \ref{toricl=2} and  \ref{toricl=3} .}, i.e.      
\begin{equation} 
\Delta_l={\rm Conv}\left(\lbrace \pm e_{i} \rbrace_{i=1}^{l}\cup \lbrace \pm (e_{i}-e_{j}) \rbrace_{1\leq i<j\leq l} \right)\ . 
\label{Delta}
\end{equation}  
Note that $\Delta_l$ contains beside these vertices no further integral point other then the origen 
$\nu_0=(0,\ldots,0)$.  Moreover, $\Delta_l$ is integral and reflexive, which implies that the dual polytope $\hat \Delta_l\subset \hat \Lambda_{\mathbb{R}}$ 
\begin{equation}
 \hat \Delta_l=\{ y\in \hat \Lambda_\mathbb{R}| \langle y,x\rangle \ge -1, \forall\  x \ \in \Delta_l\} 
 \end{equation} 
 is also an integral lattice polyhedron. Note that ${\widehat{\hat \Delta_l}}=\Delta_l$ and concretely $\hat \Delta_l$ is   
 given by
\begin{equation} 
\hat \Delta_l={\rm Conv}\left(\bigcup_{k=1}^l \bigcup_{r=1}^{\left( l\atop k\right)} \sum_{i=1}^l I_i^{(k),r} \hat e_i \cup 
\bigcup_{k=1}^l \bigcup_{r=1}^{\left( l\atop k\right)} \sum_{i=1}^l (- I_i^{(k),r} \hat e_i )\right)\ , 
\label{Delta*} 
\end{equation}  
where $\hat e_i$ is a  basis of the lattice $\hat \Lambda_\mathbb{R}$ and the $I^{(k),r}$ $r=1,\ldots, \left( l\atop k\right)$ are the sets of all 
distinct permutations of $k$ ones and $l-k$ zeros. Indeed the $2( 2^l-1)$ points listed in (\ref{Delta*}) 
are all integral points of $\hat \Delta_l$ beside the origin. For the polytope $\Delta_l$ itself it means that it has $2(2^l-1)$ faces. From the structure of the vertices of $\Delta_l$ it can 
be proven that there is no integral point in the facets of the dual polytope. The combinatorics of all facets of $\hat \Delta$ are 
equal, in particular they all have $2^{l-1}$ vertices. 

A central theorem in the  toric mirror  construction of Batyrev~\cite{MR1269718} says that  a 
smooth resolution $M$ of $M_s$ with trivial canonical class is given by the constraint 
\begin{equation}
P_{\Delta_l}= \frac{P_l(a;x)}{\prod_{i=0}^p x_i}=\sum_{\nu^{(i)}\in \Delta_l} a_i \prod_{\hat \nu^{(k)}\in \hat \Delta_l} x_k^{\langle \nu_i, \hat \nu_k\rangle} =0
\label{PDelta}
\end{equation} 
in the coordinate  ring $x_i$ of  $\mathbb{P}_{\hat \Delta_l}$, where $\nu_i$, $i=1,\ldots, p$   and   $\hat \nu_k$, $i=1,\ldots, \hat p$   
run over all integer points in $\Delta_l$ and $\hat \Delta_l$ respectively\footnote{$ P_{\Delta_l}$ is a Laurent polynomial in which the 
minimal  degree of the $x_i$ is $-1$, while $P_l(a;x)=0$ is a polynomial constraint, which  also defines  a smooth manifold 
in the  coordinate  ring .}. Here $I(\Delta_l)$ is the number of lattice points in $\Delta_l$ and  $p=I(\Delta_l)-1$. Analogous definitions  
apply for $\hat \Delta_l$. Note that (\ref{PDelta}) defines an embedding of the physical parameters $t$ and $\xi_i$, 
$i=1,\ldots, l+1$  into  convenient but  redundant complex structure variables $a_i \in \mathbb{C}$, $i=0,\ldots, l-1$. 
Both the physical as well as the  $a_i$  parameters  are only defined up to scale. Note that we are a little 
cavaliar with the notations:  The coordinate rings $x_i$, $i=1,\ldots,l+1$ in the  definition (\ref{bananageneral})  
and the one $x_i$, $i=1,\ldots, \hat p$ in   (\ref{PDelta}) are of course different. However, we can get the former 
by blowing down the latter. This is achieved by setting a suitable subset of  $\hat p -(l+1)$  of the latter $x_i$ variables to one. Likewise 
given  $P_{\Delta_l}$ in $x_i$, $i=1,\ldots, l+1$ as in (\ref{Newtonpolyhedron}) and all $\mathbb{C}^*$ action   
(\ref{C*actions}) we can uniquely extend it to $\hat p$ variables  $x_i$ by requiring that the extended 
polynomial (or  strictly speaking the proper transform of  (\ref{Newtonpolyhedron}))  is homogeneous, w.r.t. to 
all  $\mathbb{C}^*$ rescalings in (\ref{C*actions}).

The  space $\mathbb{P}_{\Delta_l}$ is a $l$-dimensional projective toric 
variety that can be  associated to any reflexive lattice polyhedra $\Delta_l$ given a star 
triangulation~\footnote{In a star triangulation all $l$-dimensional simplices  
of  the triangulation covering  the reflexive polyhedron  share the inner point, as 
one of their vertices.}  ${\cal T} $of $\Delta_l$ as 
\begin{equation}
\mathbb{P}_{\Delta_l}=\frac{\mathbb{C}^{p}[x_1,\ldots, x_p]\setminus Z_{\cal T}}{(\mathbb{C}^*)^{p-l}} \ .
\end{equation}
Here  the  $\mathbb{C}^*$ actions that are divided out are generated by 
\begin{equation}
x_i \mapsto x_i (\mu^{(k)})^{l^{(k)}_i}, \quad {\rm for} \ i=1, \ldots, p \ ,
\label{C*actions}  
\end{equation} 
where $\mu^{(k)}\in \mathbb{C}^*$ and the $l^{(k)}$ vectors span the $(p-l)$-dimensional  space of all linear relations
\begin{align}
L=\lbrace (l^*_{0},l^*_{1},...,l^*_{p})\in \mathbb{Z}^{p+1}|l^*_{0}\bar \nu_0+l^*_{1}\bar \nu_1+...+l^*_{p}\bar \nu_p=0 \rbrace
\label{defL}
\end{align}
among the  points
\begin{align}
\mathcal{A} = \lbrace\bar{\nu}_{0},\bar{\nu}_{1},...,\bar{\nu}_{p}| \bar{\nu}_{i}=(1,\nu_{i}), \nu_{i}\in \Delta_l \cap  \mathbb{Z}^{l} \rbrace \ .
\label{defA}
\end{align} 
The triangulation~\footnote{$\Delta_l$ defines canonically a fan $\Sigma_{\Delta_l}$ and the definition 
of  a smooth $\mathbb{P}_{\Delta_l}$ may require to add integer points outside $\Delta_l$ and to 
triangulate the fan  $\Sigma_{\Delta_l}$. Such cases  are discussed in~\cite{Hosono:1993qy,Hosono:1995bm}.}          
${\cal T}$ determine the set of generators  $l^{(k)}$  of $L$ and the Stanley-Reisner ideal $Z_{\cal T}$.  The 
latter describes loci in $\mathbb{C}^p[x_1,\ldots,x_p]$, which have to be excluded so that the 
orbits of the $\mathbb{C}^*$ action (\ref{C*actions}) have a well defined dimension.  
Positive linear  combinations of  $l^{(k)}$, $k=1,\ldots,n$  span the Mori cone, which is not 
necessary simplicial if $n> p-l$. It is  dual to the K\"ahler cone of  $\mathbb{P}_{\Delta_l}$ and 
all cones corresponding to all triangulations  ${\cal T}$ of  $\Sigma_{\Delta_l}$ form the secondary 
fan, see \cite{MR2810322} for a review how to calculate the $l^{(k)}$ vectors and the Stanley-Reisner ideal 
combinatorial from a triangulation ${\cal  T}$.  This combinatorics is implemented in the computer 
package \texttt{SageMath}~\cite{Sagetoric}, which calculates the possible triangulations ${\cal T}$ and from them the 
generators $l^{(k)}$ and the Stanley Reisner ideal  $Z_{\cal T}$.

The Calabi-Yau $(l-1)$- fold family defined as  section of the canonical  bundle 
$P_{\hat \Delta_l}=0$ of  $\mathbb{P}_{\Delta_l}$ is by Batryrev~\cite{MR1269718}  
conjectured to be the mirror  manifold $W=X_{\hat \Delta_l}$ of the manifold $M$, i.e. $(M,W)$ form 
a mirror pair with dual properties. A main implication of this proposal is 
that the complex structure deformation  space of $M$ denoted by ${\cal M}_{CS}(M)$ 
is identified  with the complexified  K\"ahler or stringy K\"ahler moduli space 
${\cal M}_{KCS}(W)$
\begin{equation}
{\cal M}_{KCS}(W)={\cal M}_{CS}(M)
\label{eq:identification}
\end{equation} 
and vice versa. Note that the real K\"ahler moduli space is parametrized by the 
K\"ahler parameters $t^{\mathbb{R}}_k=\int_{{\cal C}_k} \omega$, where $\omega\in H^{1,1}(M)$ and 
${\cal C}_k$ span a basis of holomorphic curves in $H_{1,1}(M,\mathbb{Z})$.  In string theory the 
complexification is due to the Neveu-Schwarz  two-form field $b$  also in $H^{1,1}(M)$.  
The complex  variables $t_k=\int_{{\cal C}_k} \omega+ i b$, $k=1,\ldots,h_{1,1}(M)$  
parametrize locally the complexified  moduli  space ${\cal M}_{KCS}(W)$ of $W$.

We will next discuss  the space  ${\cal M}_{cs}(M)$ of complex structure 
deformations  of $M$.  This space is redundantly parametrized  by the  
complex coefficients $a_i$, $i=0,\ldots, l(\Delta_l)-1$ in (\ref{PDelta}). 
The $a_i$ are identified by $l+1$ scaling 
relations on the coordinates of  $\mathbb{P}_{\hat\Delta_l}$ and the automorphism of  $\mathbb{P}_{\hat\Delta_l}$ 
that leaves $M$ invariant but acts on the parametrizations of the polynomial constraint $P(a;x)$.  The latter 
one parameter families of identifications of the deformation parameters  are in an one-to-one 
correspondence to the points inside codimension one faces of $\Delta_l$.  Let us denote  by $\Theta^j_k$ all faces of codimension $k$ in 
$\Delta_l$ labeled by  $j$.  $I(\Theta^j_k)$  denotes the number  of lattice points contained  
in $\Theta^j_k$, while $I'(\Theta^j_k)$ denotes the number of lattice points that 
lie in the interior of $\Theta^{(j)}_k$. With this notation  $M$ has  
$I(\Delta_l)-(l+1) -\sum_{j} I'(\Theta^j_1)$   independent complex structure deformations. 
They  correspond  to elements in $H^1(M,TM)$  and are unobstructed on a Calabi-Yau  
manifold $M$. The cohomology group $H^1(M,TM)$ is related to the cohomology group 
$H^{l-2,1}(M)$ via the contraction with the unique holomorphic $(l-1,0)$-form.

Equation (\ref{eq:identification})  implies that in particular the complex dimensions of these spaces have to 
match, i.e.  $h_{1,1} (X_{\hat \Delta_l})=h_{l-2,1}(X_{\Delta_l})$ and $h_{1,1} (X_{\Delta_l})=h_{l-2,1}(X_{\hat \Delta_l})$.  
From theses facts it follows that the dimensions of these important  cohomology groups are given by counting 
integral points in the polytops\footnote{The last terms  after the first equal sign in the formulas in each 
line of (\ref{Hodge}) correspond to K\"ahler--- or complex structure deformations, which are frozen by  
the toric realization of the manifolds, respectively. Likewise the  third terms are absent in our case. The last 
equality holds only for the polyhedra given in (\ref{Delta}) and (\ref{Delta*}).}               
\begin{gleichung}
& h_{1,1}(X_{\Delta_l})=I(\hat \Delta_l)-(l+1)- \sum_{j} I'(\hat \Theta^j_1)+\sum_{j} I'(\hat \Theta^i_2) I'(\Theta^i_{l-2}))  =2^{l+1}-l-2\\
& h_{l-2,1}(X_{\Delta_l})=I(\Delta_l)-(l+1)- \sum_{i} I'(\Theta^{i}_1)+\sum_{i} I'(\Theta^i_2) I'(\hat \Theta^i_{l-2})) =l^2\ . 
\label{Hodge}
\end{gleichung}      

For $l=3$ the Calabi-Yau manifold $M$ will be a nine-parameter family  of polarized $K3$ surfaces. In this case the transversal 
cycles in $h_{11}$ are counted $h^T_{11}=I(\Delta_l)-(l+1)=9$, i.e. in total one has eleven transcendental and  
eleven algebraic two-cycles, which are counted by $h^A_{11}=I(\hat \Delta_l)-(l+1)=11$. For $l=4$ the $16$-parameter family of 
Calabi-Yau three-fold has $h_{11}=26$   and $h_{21}=16$ and hence Euler 
number $\chi=40$. For $l=5$ the Calabi-Yau four-fold has $h_{31}=25$, $h_{11}=57$, $h_{21}=0$ and $\chi=540$. Using 
an index theorem~\cite{Klemm:1996ts} one gets $h_{22}=422$.   

Since our polytope  (\ref{Delta})  has only $\sum_i I(\Theta^{i}_l)=l(l+1)$ corners and  one inner points the manifold 
$M$ has $l^2$ complex structure deformations, which have to be eventually mapped to our physical parameters 
$t$ and $\xi_i$. Since the latter are equivalent up to scaling by $\mu$ we have  $l+1$ independent physical 
parameters. Therefore, the map to the physical parameter space has a huge kernel for high $l$ and  special effort has to be 
made to specify the relevant physical  subspace of ${\cal M}_{phys}(M) \subset  {\cal M}_{cs}(M)$ as described  concretely 
in the example sections \ref{1looptex}, \ref{sunsetex} and \ref{3looptex}.

Actual properties of the smooth canonical  resolution of $M_s$, in particular its K\"ahler cone, 
depend on  the choice of the star  triangulation $\hat {\cal T}$  of $\hat \Delta_l$. However, 
these detailed  properties of the K\"ahler moduli space ${\cal M}_{KS}(M)$ of $M$ do not affect 
the complex  moduli space  ${\cal M}_{CS}(M)$ and the integral (\ref{bananageneral})  over closed cycles, 
like  ${\cal F}_{T^{l}}$, the  integral over the $T^{l}$ torus. This maximal cut integral depends only 
on the complex structure parameters.  The blow up coordinate  ring allows however a useful description of the 
boundary contribution to ${\cal F}_{\sigma_l}$, see~\cite{Bloch:2016izu}.  Moreover, the identification  
(\ref{eq:identification}) turns out to be very  useful to introduce suitable coordinates on  ${\cal M}_{CS}(M)$ to obtain solutions for 
the integral (\ref{bananageneral}). Different star triangulations ${\cal  T}$ of the polyhedra $\Delta_l$  correspond to different 
K\"ahler cones of the ambient space $\mathbb{P}_{\Delta_l}$ of $W$ and correspond eventually\footnote{If all  curves 
that bound the K\"ahler cone of  $\mathbb{P}_{\Delta_l}$ descend to the hypersurface $W$.}  
to different K\"ahler cones of $W$. Each choice of the K\"ahler cone of $W$, 
defines by mirror symmetry and the identification (\ref{eq:identification}) canonical so called Batyrev coordinates $z_i$, 
$i=1,\ldots, h_{l-2,1}(M)=h_{1,1}(W)$  on  ${\cal M}_{CS}(M)$, at  whose origin $z_i=0$ for all $i$ 
there is  a point of maximal unipotent monodromy in ${\cal M}_{CS}(M)$. The coordinates $z_i$ are 
ratios of the coefficients $a_i$ of $P_{\Delta_l}$ given for each triangulation by
\begin{equation} 
z_k= (-a_0)^{l^{(k)}_0} \prod_i a_i^{l^{(k)}_i}, \quad  k=1,\ldots, p-l\ . 
\label{zcoords}
\end{equation}  
The definition of the $z_k$  eliminates the scaling relation.  Since in our case we have 
no codimension one points, i.e. no automorphism  of $\mathbb{P}_{\hat \Delta}$ leaving  
the hypersurface  invariant and further  identifying the $a_i$ deformations, the $z_k$ 
are actual coordinates on ${\cal M}_{CS}(M)$. In other simple cases one can restrict in the definition of $L$ (\ref{defL}) to linear relations of points, which are 
not in codimensions one, the general case is discussed in \cite{Hosono:1993qy,Hosono:1995bm}.  In the moduli space of ${\cal M}_{CS}(M)$  as parametrized  by the independent 
$a_i$, the $z_k$ are blow up coordinates  resolving  singular loci in discriminant components of  
the hypersurface $P_{\Delta_l}=0$ in ${\cal M}_{CS}(M)$, so that these become in the resolved model of the complex moduli 
space  $\widehat{ {\cal M}_{CS}}(M)$ intersection points  of normal crossing divisors $D_i=\{z_i=0\}$, $i=1,\ldots, h_{l-2,1}(M)$.                    

Of particular significance in the geometric toric construction of the 
differentials on $M$ is the coefficient  $a_0$ of the monomial $\prod_{i}^{l+1} x_i$ in  $P(t,\xi;x)$ 
corresponding to the 
inner point in $\Delta_l$, which is given in the physical parameters by 
\begin{equation}
u\coloneqq a_0=  t- \sum_{i=1}^{l+1} \xi_i^2\ . 
\label{coefficientinnerpoint}
\end{equation}               

The families that are just parametrized by $u$ with the coefficients of all other points set to one, i.e. 
in particular $\xi^2=1$ for all $i=1,\dots, l+1$ is particularly symmetric. For $l=4$, i.e. Calabi-Yau three-folds,   the family is known as the  
Barth-Nieto quintic. The form of this family is conveniently given by a complete  intersection in $\mathbb{P}^{l+1}$ that can be readily 
generalized to the ones 
\begin{equation} 
\sum_{i=1}^{l+2} x_i=0 \quad\text{and}\quad \sum_{i=1}^{l+1} \frac{1}{x_i}+ \frac{1}{u x_{l+2}}=0\ .
\label{BarthNieto}
\end{equation}
By solving for $x_{l+2}$  and homogenizing one  gets $P_{\Delta_l}=0$ in the equal mass case parametrized by 
$u$ and for equal masses  $\xi^2_i=1$ for all $i=1,\dots, l+1$. 

\subsubsection{Period integrals on $M$ and maximal cut amplitude}
\label{periodsgkz}

For the discussion of the period integrals, which are very close to the integral of interest (\ref{bananageneral}),     
we start with a residue definition of the holomorphic $(n,0)$-form $\Omega$  of the Calabi-Yau manifold $M$ of 
complex dimension $n=l-1$ defined as hypersurface in a toric ambient space
\begin{equation} 
\Omega= \oint_\gamma  \frac{ a_0 \mu_l }{P_l(a;x)}~,   
\label{Omega}
\end{equation}  
where $\gamma$ encircles the locus $P_l=0$ in the toric ambient  space and $\mu_l$ was defined in \eqref{measure}.
Given a basis $\Gamma_i$ of the cycles in the middle dimensional homology $H_n(M,\mathbb{Z})$  we can define 
closed string period integrals
\begin{equation} 
\Pi(\Gamma)=\int_{\Gamma} \Omega \ . 
\label{Periodcycle}
\end{equation} 
The closed string periods are directly relevant as one of them  describes the maximal cut integral. 
Moreover, by the local Torelli theorem $h^{n-1,1}$ of them can serve as projective coordinates of 
${\cal M}_{CS}$ and by Griffiths transversality the periods fulfill differential relations for odd $n>1$, 
algebraic relations for $n=2$ and algebraic as well as differential relations for even $n>2$.

At the point of maximal unipotent monodromy  that is specified   as the origin of the Batyrev coordinates $z_i$, $i=1,\ldots h$ from (\ref{zcoords}), which are  simply  defined by the Mori cone $l^{(k)}$ vectors of the mirror $W$, 
the Picard-Fuchs differential ideal is maximally degenerate. This point is a point of maximal unipotent
monodromy or short MUM-point.   As a consequence  that near the MUM-point 
there is exactly one holomorphic period, and for $k=1,\ldots, n$ there are $h_{hor}^{n-k,k}$ periods 
whose leading multi-degree in $\log(z_i)$, $i=1,\dots, h^{n-1,1}$  is of order $k$.  For Calabi-Yau $n$-folds 
with $n>2$ the  full cohomology groups $H^{n,0}, H^{n-1,1}$  are horizontal. By complex conjugation this holds also for 
$H^{1,n-1}$ and $H^{0,n}$. In particular, for Calabi-Yau three-folds the whole middle cohomology is horizontal.
Beside this general structure an additional  bonus in the case of   Calabi-Yau spaces 
given  by hypersurfaces in  toric ambient spaces is that there is a $n$-cycle with the topology 
of a $n$-torus $T^n\in H_n (M,\mathbb{Z})$  which yields that holomorphic period $\varpi\coloneqq\Pi(T^n)$ explicitly.   
With the definitions (\ref{Omega}) and (\ref{Periodcycle})  this integral  yields an $(n+1)$-times 
iterated residue integral over an $T^{n+1}$ in the ambient space, that can be  readily evaluated in terms of the $l^{(k)}$ 
vectors as 
\begin{equation}
\varpi=\oint_{|x_1|=0} \frac{d x_1}{2 \pi i}   \ldots \oint_{|x_{n+1}|=0} \frac{d x_{n+1}}{2 \pi i}  \frac{a_0}{P_l(a;x)} =
\sum_{\{\underline k\}} \frac{ \Gamma\left( - \sum_{\alpha=1}^h l^{(\alpha)}_0 k_\alpha+1\right)}{\prod_{l=1}^p  \Gamma\left(\sum_{\alpha=1}^h l^{(\alpha)}_l k_\alpha +1\right)}
\prod_{\alpha=1}^h z_\alpha ^{k_\alpha}   \ . 
\label{varpi}  
\end{equation}     
Here we use the coordinate ring $x$  as in  (\ref{bananageneral}) and set $x_{n+2}=1$. 
In the tuple $\{ {\underline k}\}=\{k_1,\ldots , k_h\}$ each $k_i$ runs over non negative 
integers $k_i \in \mathbb{N}_0$ and  $p$ is defined in (\ref{defL}).  Note that  
by de\-fi\-ni\-tion the sum of the  integer entries in each $l^{(k)}$ is zero, therefore they have  negative entries. 
For hypersurfaces and  complete intersections the $l_0^{(k)}$ entry is non-positive   
$l_0^{(k)}\le 0 $ for all $k$. However, for  $i>0$ the $l_i^{(k)}$ can have either sign. Poles 
of the  $\Gamma$-function at negative integers in the denominator make the 
summand vanishing. 
This effectively restricts the range of  the $\{k_1,\ldots , k_h\}$ to a positive cone  
\begin{gleichung}
	\sum_{\alpha=1}^h l_j^{(\alpha)} k_\alpha \geq 0 \ .
	\label{rangesum} 
\end{gleichung}
Restricting  to the physical slice, i.e. to $\underline z(t,\xi_i)$ of the $l$-loop graph, means to parametrize the 
$a_i$, $i=0, \hdots , h=l^2$ by the physical variables $t,\xi_i$. Due to the  definition of 
(\ref{zcoords}) one can find a splitting of the set of indices  $\{\alpha_1,\ldots, \alpha_{h}\}$ 
into $\{ \alpha_1,\ldots, \alpha_{l+1}\}$ and  $\{\alpha_{l+2},\ldots, \alpha_{h}\}$  so that the 
variables $\{z_{\alpha_{l+2}},\ldots, z_{\alpha_{h}}\}$ are  either set to constant values or 
identified with the variables $z_{\alpha_j}(t,\xi)$, $i=1,\ldots, l+1$.  A key observation in 
the examples is that the range  (\ref{rangesum}) is such that the  contribution from the 
summation over the $k_{\alpha_j}$, $j=l+2,\ldots h$  to each monomial  $\prod_{i=1}^{l+1} z_{\alpha_i}^{k_i}$ is finite.  
This implies in that (\ref{varpi}) can also be given non-redundantly in $l+1$ physical parameters 
$z_{\alpha_j}(t,\xi)$, $i=1,\ldots, l+1$ exactly to arbitrary order.  The range (\ref{rangesum})  and (\ref{varpi}) can also be calculated  directly  
as follows: Expanding in the integrand  $a_0/P_l(x,a)=\left[1/\prod_i x_i\right] \left[1/(1-1/a_0(\ldots))\right]$  the 
second factor as a geometric series and  noticing that only the constant terms of it  
contribute to the integral yields the result.  Applying this  to the $P_l$  in 
(\ref{bananageneral}) yields the all $(l=n+1)$-loop maximal cut  integrals 
\begin{equation}
 {\cal F}_{T^{l}}(t,\xi_{i}) =  \frac{ \varpi\left(\underline z(t,\xi_i)\right)}{t- \sum_{i=1}^{l+1} \xi_i^2}
 \end{equation} 
as an exact series expansion with finite  radius of convergence for regions in the 
physical parameters in which  $z_k(t,\xi_i)$ are all small.

In principal, one can analytically continue this to all regions in the physical parameter space. 
This task can greatly aided if one  knows  the Picard differential ideal that annihilates $\varpi$ 
and  all other periods.  The derivation of the latter  will be discussed  in the next section. It certainly
helps if one knows all other periods   near $z_k=0$.  Because of the structure of the logarithmic 
solutions at the MUM-point these can by easily given by the Frobenius method. This is done by introducing $h$ auxiliary 
deformation parameters $\rho_\alpha$ in 
\begin{equation} 
\varpi({\underline{z}},{\underline{\rho}})=\sum_{\{\underline k\}} c({\underline k}, {\underline \rho})  {\underline z}^{\underline k+\underline\rho} ,       
\label{frobsol}
\end{equation} 
where  ${\underline z}^{\underline k+\underline\rho}:= \prod_{\alpha=1}^h z_\alpha ^{k_\alpha+\rho_\alpha}$ and 
\begin{equation}
c(\underline k, \underline \rho)= \frac{ \Gamma\left( - \sum_{\alpha=1}^h l^{(\alpha)}_0 (k_\alpha+\rho_\alpha)+1\right)}{\prod_{l=1}^p  \Gamma\left(\sum_{\alpha=1}^h l^{(\alpha)}_l (k_\alpha+\rho_\alpha +1\right)}    \ .
\end{equation} 
With this definition $\varpi({\underline z})= \varpi({\underline{z}},{\underline{\rho}})|_{{\underline \rho}=0}$ the $h^{n-1,1}$ 
linear logarithmic solutions are given by 
\begin{equation}  
\Pi(\Gamma_\alpha)= [(1/( 2 \pi i) \partial_{\rho_{\alpha}}  \varpi({\underline{z}},{\underline{\rho}})]|_{{\underline \rho}=0}=1/(2 \pi i) \Pi(T^n) \log(z_\alpha) +{\cal O}(z) \ . 
 \label{logsol}
\end{equation} 
It can be shown that  $\Gamma_\alpha\in H_m(M,\mathbb{Z})$.  All other solutions corresponding to  the rest of the cycles $\Gamma_\beta\in H_n(M,\mathbb{Z}) $ 
are of order $2 \le k \le n$ in the logarithms and of the form  
\begin{equation} 
\Pi(\Gamma_\beta)= [c_\beta^{\alpha_1\ldots \alpha_k}  \partial_{\rho_{\alpha_1}} \ldots   \partial_{\rho_{\alpha_k}} \varpi({\underline{z}},{\underline{\rho}})]|_{{\underline \rho}=0} \ ,
\label{higherlogsol}
\end{equation}  
where the tensors  $c_\beta^{\alpha_1\ldots \alpha_k}$ contain 
transcendental numbers fixed by the $\hat \Gamma$-class conjecture and classical intersection theory on $W$, 
see~\cite{MR3965409} for a review.

 \subsubsection{GKZ systems and Picard Fuchs differential ideal}    
Gel\cprime fand, Kapranov and Zelevinsk\u{\i}~\cite{MR1020882} investigated integrals of the from 
\begin{equation} 
{\cal F}^{GKZ}_\sigma=\int_{\sigma} \prod_{i=1}^r P(x_1,\ldots, x_k)^{\alpha_i} x_1^{\beta_1}\cdots x_k^{\beta_k} \mathrm d x_1\cdots \mathrm d x_k \ ,
\label{GKZ}       
\end{equation} 
which can be specialized  to (\ref{bananageneral}), which is in turn similiar to (\ref{varpi}), even though in 
(\ref{varpi}) we took the integration domain  to be a closed cycle $T^{n+1}$,  while~\cite{MR1020882} just 
speek of cycles $\sigma$. 

In  (\ref{bananageneral}) $\sigma$  is a closed cycle only for the maximal cut case which leads to 
(\ref{varpi}), otherwise $\sigma$ is  a chain.  In this case the corresponding differential ideal, which 
is fulfilled by the integral (\ref{GKZ})   is inhomogeneous. The GKZ integrals can be viewed as 
systematic multivariable generalization  of the Euler integral $_2F_1(a,b,c;z)=\sum_{n=0}^\infty\frac{(a)_n (b)_n}{n! (c)_n} 
=\frac{\Gamma(c)}{\Gamma(b) \Gamma(b-c)}  \int_0^1 t^{(b-1)}(1- t)^{(b-c-1)} ( 1- z t)^{-a}$, which  solves  
Gauss hypergeometric systems and $\varpi$ as a specially simple  generalized multivariable  hypergeometric series.   

As mentioned  at the end of the introduction to subsection (\ref{ssec:geometry}) at least for integer exponents  
the requirement that these higher dimensional integrals are well defined under the sca\-ling symmetries of 
the parameters, that appear in physical Feynman integrals,  is equivalent to the vanishing of the first Chern 
class and hence  these Feynman integrals with $r=1$, $\alpha_1=-1=-n_1$ are closely related to 
period integrals over the holomorphic $(n,0)$-form in the cohomology group $H^{n,0}$ of the Calabi-Yau manifolds $M$  defined as 
hypersurfaces in toric varieties~\cite{Hosono:1993qy,Batyrev:1993wa,Hosono:1995bm}. The same argument  
relates integrals with $r>1$ and $\alpha_i=-1=-n_i$ to complete intersection Calabi-Yau  spaces~\cite{Hosono:1994ax,MR1328251,MR1463173}. 

More general integrals are  related to the former by taking derivatives w.r.t. to the independent complex moduli parameters 
say $a$. In particular, such derivatives change the Hodge type of the integrand as follows. Let $F^p(M)=\bigoplus_{l\ge p} H^{l,n-l}(M)$ a Hodge 
filtration $H^n=F^0\supset F^1\supset \ldots \supset (F^n=H^{n,0})\supset F^{n+1}=0$, then   $H^{p,q}(M)=F^p(M)\cap {\overline{ F^q(M)}}$, 
and the $F^p(M)$ can be extended to holomorphic bundles ${\cal F}^p(M)$  over the complex 
family $M$ over ${\cal M}_{CS}(M)$, with 
\begin{equation}
\partial^k_{a} {\cal F}^{n}(M) \in {\cal F}^{n-k}(M) \ . 
\label{transversal} 
\end{equation}    
Since  the bundles  ${\cal F}^p(M)$ are of finite rank, there will be differential  relations among finite 
derivatives w.r.t. to the moduli, which implies  that the period integrals  over closed cycles are 
annihilated by finite order linear differential operators  ${\cal D}_k$, where the derivations are w.r.t.  
the moduli and the coefficients are rational functions in the moduli. In particular, one  can specify a differential  
ideal, called the Picard-Fuchs differential ideal,   ${\cal D}_k$,  $k=1,\ldots, d$ that determines the 
periods as finite linear combination of its system of solutions.

One key tool to find the  differential relations between these integrals is the Griffiths  reduction 
method, which relies on the following partial integration formula, that is valid up to exact terms, 
i.e. holds under the integration over closed  cycles~\cite{MR0260733}
\begin{equation}
\sum_{k\ne j}{n_k\over n_j-1}{P_j\over P_k}
{Q \partial_{x_i} P_k\over \prod_{l=1}^r P_l^{n_l}}\mu\ =
{1\over n_j-1} {P_j \partial_{x_i} Q\over \prod_{l=1}^r P_l^{n_l}}\mu -{Q \partial_{x_i} P_j\over \prod_{l=1}^r P_l^{n_l}}\mu \ ,
\label{partialintegrationcicy}
\end{equation}
where $Q(x)$ are polynomials of the appropriate degree to ensure the scale invariances and $\mu$ 
is straightforward generalization of the measure (\ref{measure}).  Such $Q(x)$  
arise automatically, when partial derivative  w.r.t. the moduli are taken.      
Using these equations and  Gr\"ober basis calculus one can reduce higher 
derivatives w.r.t. to the moduli to lower ones and find eventually the complete 
differential ideal. These relations between rational functions are also used in the literature not only to compute differential equations for Feynman integrals
but also for finding so called master integrals. If these master integrals are known with the partial integration relations \eqref{partialintegrationcicy} the whole Feynman integral is evaluated. For a review on master integrals in Feynman graph computation we refer to \cite{Henn:2014qga}.

However,  this method is computationally very expensive in multi moduli cases. 
Therefore, we employ as far as possible a different derivation of differential 
relations which follow from scaling symmetries  that follow from the combinatorics 
of the Newton polytope, known as GKZ differential  system.   For this purpose we 
define    
\begin{equation} 
\hat \Omega= \oint_\gamma  \frac{\mu_l }{P_l(a;x)} \quad\text{and}\quad  \hat \Pi_\sigma=\int_{\sigma}  \frac{\mu_l }{P_l(a;x)}~.
\label{hatPi}
\end{equation} 
          
Now each linear relation among the points in the Newton polytope as expressed by the $l^{(k)}$-vectors, $k=1,\ldots,l^2$ yields a 
differential operator $\mathcal{D}_{l^{(k)}}$ in the redundant moduli $a$. Moreover, the  infinitesimal invariance  under 
the $(\mathbb{C}^*)^{n+2}$ scaling relations yields further dif\-fer\-en\-tial operators  $\mathcal{Z}_{j}$, $j=1,\ldots,n+2$. Together they  
constitute an resonant GKZ system~\cite{MR1011353,MR1080980}: 
\begin{align}
&{\hat {\mathcal{D}}}_{l^{(k)}} {\hat \Pi}_\sigma= \left(\prod_{l^{(k)}_{i}>0}\left(\dfrac{\partial}{\partial a_{i}}\right)^{l^{(k)}_{i}}-\prod_{l^{(k)}_{i}<0}\left(\dfrac{\partial}{\partial a_{i}}\right)^{-l^{(k)}_{i}}\right) {\hat \Pi}_\sigma=0 \quad\text{and}	\\
&\mathcal{Z}_{j} {\hat \Pi}_\sigma=\left(\sum_{i=0}^{p} \bar{\nu}_{i,j}\theta_{a_{i}}-\beta_{j}\right) {\hat  \Pi}_\sigma =0
\label{gkzdef}
\end{align}
with $\beta=(-1,0,...,0)\in \mathbb{R}^{n+2}$ for the hypersurface case and $\theta_{a}=a\partial_{a}$, in the form that 
applies to the integrals  in Calabi-Yau  hypersurfaces in toric varieties ~\cite{MR1269718,Hosono:1993qy}, for 
which the integration domain $\sigma$ is also scale invariant.  In this case we can  use the relations 
$\mathcal{Z}_{j} \hat \Pi_\sigma=0$ to eliminate the $a_i$ in favour of the scale invariant $z_i$ defined in  (\ref{zcoords}) 
using $a_i \partial_{a_i} = \sum_{k=1}^{l^2} l^{(k)}_i z_k \partial_{z_k}$ and by the commutation relation $[ \theta_{a}, a^r]=r a^r $  
applied previously  to $a_0$ we obtain operators  $\mathcal{D}_{l^{(k)}}(z)$ that annihilate $\Pi(\Gamma)$. As it turns 
out these operators do not determine the  $\Pi(\Gamma)$ as they admit further solutions~\cite{Hosono:1993qy}. To obtain 
the actual Picard-Fuchs differential ideal one can factorize the $\mathcal{D}_{l^{(k)}}(z)$  and disregard trivial factors that allow
for additional solutions which have the wrong asymptotic to be periods~\cite{Hosono:1993qy,Hosono:1994ax}.  
In practice the most efficient  way to get the Picard-Fuchs differential ideal is often to make an ansatz  for additional  
minimal order differential  operators that annihilate (\ref{varpi}) and check that the total  system of differential  
operators allows no additional solutions then the ones specified in \eqref{logsol} and \eqref{higherlogsol}.

One of our main results is that we give the general strategy to derive the Picard-Fuchs differential ideal in the
physical parameters  $z_i(t,\xi)$, $i=1,\ldots,l+1$ and give it explicitly  for one, two and three loops in 
equations (\ref{opbubble}), (\ref{opsunset}) and in (A.1)-(A.4) for the three-loop banana graph. These 
operators determine the maximal cut integral everywhere in the parameter space. By applying these operators 
to the geometrical chain integral 
\begin{equation} 
\Pi_{\sigma_l}=\int_{\sigma_l}  \frac{a_0 \mu_l }{P_l(a;x)}
\label{opengeometric} 
\end{equation} 
and integrating explicitly over the boundary of the chain we can find the 
inhomogeneous  differential equations and the corresponding special solutions 
describing the full $l$-loop banana graphs explicitly up to three  loops.  

Let us end this section with some remarks on additional structures  for the periods of 
Calabi-Yau $n$-folds, which are relevant to understand   the differential ideal that determines 
the maximal cut integral  better. For a given basis of transcendental $n$-cycles $\Gamma_i\in H_n(M,\mathbb{Z})$ one 
can find dual elements $\gamma_j\in H^n_{hor}(M,\mathbb{C})$ so that 
$\int_{\Gamma_i} \gamma^j=\delta_i^j$ and expand the holomorphic $(n,0)$-form $\Omega= \sum_i \Pi(\Gamma_i) \gamma^i$. 
Let us define for each set $A$  of indices of order $r$  the order $r$ differential  operator 
$\partial^r_A :=\partial_{z_{a_1}} \ldots \partial_{z_{a_r}}$. Then by (\ref{transversal}) and consideration 
of type one gets the {\em transversality conditions}~\cite{MR717607}  
\begin{equation}
\label{specialI}
\int_M \Omega \wedge  \partial^r_A \Omega \; = \;  \Pi(\Gamma_i) \Sigma^{ij}  \partial^r_A  \Pi(\Gamma_j)\; = \; 
\left\{
\begin{array}{ll} 
0&\ {\rm if}\  r< n\\ [ 3 mm]
C_{A}(z) & \ {\rm if}\ r\; = \;n\ ,  \\ 
\end{array}
\right. 
\end{equation} 
where $C_{A}$ are rational functions in the $z_i$, known as Yukawa couplings for $n=3$. The 
form $\Sigma^{ij}=\int_M \gamma^i\wedge \gamma^j$ is integer and symmetric for $n$ even and antisymmetric  
for $n$ odd. In the latter case one can chose a symplectic basis for the $\gamma^i$. For the $\mathrm{K3}$ 
or more generally $n\ge 2$ and $n$ even it implies that the solutions to the Picard-Fuchs differential ideal fulfill nontrivial  quadratic relations
\begin{equation} 
\Pi(\Gamma_i) \Sigma^{ij}\Pi(\Gamma_j) =0	\quad\text{and}\quad	 \Pi(\Gamma_i) \Sigma^{ij}\partial_{z_k} \Pi(\Gamma_j) =0~, \quad \forall k \ . 
\label{quad}  
\end{equation}
We will discuss the consequences at the level of the differential operator more  in section \ref{equalmassK3}. 
For $n=3$ it implies special geometry, see \cite{MR3965409} for a review.

\subsubsection{Geometrical and physical periods}
\label{ssec:geomperiods}
The physical moduli space of the banana Feynman diagrams is parametrized by the $l+2$ parameters $(t, \xi_1, \xi_2, \hdots, \xi_{l+1})$, where additionally one of these can be scaled away. As mentioned, compared to the moduli space parametrized by all Batyrev coordinates $z_i$ the physical moduli space is much smaller. In the following we explain how one can make a restriction onto the physical moduli space. 

Besides this restriction there is another difficulty we have to mention. For the description of the large moduli space through the Batyrev coordinates $z_i$ it is crucial to have a minimal number of Mori cone generators. They are determined from the triangulation $\mathcal T$ of the polytope\footnote{For $l=3$ one can easily get all $2^{6}$ star triangulations but for $l=4$ there is an extremely 
large number of different star triangulations, which we have estimated to be  $6^{20}$. Listing all of them cannot be done by a desktop computer.} $\Delta_l$. There are only finitely many fine and star-triangulations such that it is not directly clear that there exists a triangulation which yield $l^2$ Mori cone generators. Actually, for the sunset graph this is the case. In such a situation one starts with a triangulation yielding a non-minimal number of Mori cone generators. We claim that one can still take out $l^2$ $l$-vectors describing the Feynman graph geometry appropriately. The choice of $l^2$ vectors is neither arbitrary nor unique but we can give some criteria\footnote{We do not claim that these criteria are necessary or sufficient.} for choosing them correctly. Different proper selections of $l$-vectors should at the end yield the same results for the Feynman graph.

First of all the $l^2$ vectors should be all linear independent over the real numbers. Secondly, we want $l+1$ $l$-vectors having a non-vanishing entry for the inner point which are important in the physical limit. Furthermore, we want that in the $i$-th components of all $l$-vectors there is at least a positive entry. This should be true for all components $i$ without the one for the inner point. From the last condition we hope that it guarantees that the structure of solutions is as we explained in section \ref{periodsgkz}. This one can check by analyzing that the GKZ operators defined in \eqref{gkzdef} do indeed annihilate the Frobenius solutions with positive powers \eqref{frobsol}.

We think that these conditions give a strategy to take out the required $l^2$ mori cone generators. For the sunset graph we have to follow this strategy and we give the results in section \ref{sunsetex}. Although there exist fine and star-triangulations with nine $l$-vectors for the three-loop banana diagram, we nevertheless applied our criteria on a non-simplicial cone. Also in the three-loop case the criteria select a proper set of nine $l$-vectors yielding the same results as presented in section \ref{3looptex} computed from a triangulation with minimal number of mori cone generators.

Now the restriction onto the physical moduli space starts with using the inequalities \eqref{rangesum} such that the holomorphic solution \eqref{varpi} is evaluated exactly in the physical relevant Batyrev coordinates. Having found this period on the physical slice we search for operators annihilating it such that the set of common solutions to these operators form a basis of the periods on the physical slice. This finally yields a basis of periods on the physical moduli space. It is quite hard to give a universal description of these operators. In general they form a differential operator ideal of linear, homogeneous differential operators and their explicit form as for example their degree depend on the representation of the ideal. For our discussion we write down an ansatz for a differential operator in terms of logarithmic derivatives of the remaining Batyrev coordinates. Thereby, we start with second order operators with polynomial coefficients which we make of smallest degree as possible. Typically, this ansatz yields a large number of possible operators from which we have to take a generating set of the differential ideal. From cohomology arguments we expect as many single logarithmic solutions as the number of interesting physical parameters, which strongly depends on the concrete banana diagram. Therefore, we take as many operators until their number of logarithmic solutions fits to the cohomological prediction. If the resulting solutions do still not satisfy all expectations, e.g. the number of higher logarithmic solutions, one has to extend the set of operators with higher degree ones until all expected solutions are determined. In this way one finds a generating set of operators for the differential ideal describing the physical periods. This part of our method depends strongly on the given form of the physical holomorphic period which is why we refer to our examples. We only remark that later it is crucial that the operators and the physical solutions are expressed in the remaining physical Batyrev coordinates.

\subsection{The complete banana diagram and inhomogeneous differential equations}
\label{inhomstrategy}

So far, we have found a complete differential ideal with solutions spanning a basis of the physical periods. Or said differently, these functions after dividing by the inner point describe the maximal cut integral $\mathcal F_{T^l}$. Now we extend our method to find the missing functions which complete the function space for the full banana Feynman diagram $\mathcal F_{\sigma_l}$. By function space we mean a set of functions which suitably combined yield the complete banana Feynman integral \eqref{bananageneral}. It turns out that for the banana graphs there is only a single additional function we have to compute.

Basically, we extend the homogeneous differential ideal to a set of inhomogeneous differential operators such that its solutions describe the full Feynman graph. These inhomogeneities are found from the appropriate homogeneous operator by the following process: We let an operator directly act on the geometric differential, which is given as the integrand of \eqref{opengeometric}, and perform then the integration over the domain $\sigma_l$. In this way we obtain for every homogeneous operator a corresponding inhomogeneous one.

For this task the original parametrization of the differential is changed to the Batyrev coordinates \eqref{zcoords}. This has a major advantage in the following. After applying the operators on the differential we can integrate over the simplex $\sigma_l$. In contrast to a period integral the integration range of the complete Feynman graph is not closed and such we get non zero after integration. Unfortunately, these integrals can not be carried out analytically with generic parameters. But they can be performed easily numerically. The advantage of including the inner point and using the Batyrev coordinates is now that the numerical results can simply be guessed. We claim that for the $l$-loop banana integrals they are only given as linear combinations of logarithms in the Batyrev coordinates. In our calculated examples given in section \ref{examples} we could always guess the inhomogeneities yielding a full set of inhomogeneous operators.

In the literature there are already some methods known for computing relative periods in a way that homogeneous differential equations describing usual periods are extended to inhomogeneous ones. For examples in \cite{Li:2009dz} a method for general toric varieties is explained how to extend the GKZ method to relative periods. The key point for this method is the $l$-vector description of the variety and its relative cohomology. The $l$-loop banana diagrams are not entirely described through $l$-vectors and therefore this method can not be applied. Moreover, there is the Dwork-Griffith reduction to obtain the homogeneous differential equations which then can analogously be extended to inhomogeneous ones as in our method \cite{MR3780269}. Although Dwork-Griffith reduction can in principle be applied in any situation as explained before, for computational reasons only the sunset graph can explicitly be done. Compared with known methods our strategy uses the structure of the $l$-loop banana diagrams more efficiently and produces results also for high loop orders.

Having found the inhomogeneous operators its solutions are given by the solutions of the homogeneous operators together with a single special solution of the inhomogeneous system. A special solution is found by an ansatz which has a similar logarithmic structure as the homogeneous solutions. Only the power of the highest appearing logarithm is increased by one compared to the other solutions. This closes the set of functions describing the $l$-loop banana Feynman graph.

Our method gives a relatively small set of functions necessary to compute the banana graphs. For example, with numerical computations the correct linear combination of these functions evaluating to the Feynman graph can be fixed. We exemplify this on the sunset graph in section \ref{subsecresultssunset}. Moreover, a detailed analysis of the analytic structure of these functions based on the inhomogeneous differential equations can be elaborated and produce new insights of the Feynman graph, for instance branch cuts or singularities representing particle productions.  


\section{Examples}
\label{examples}

In this chapter we explain our method by means of three different examples, the one-, two- and three-loop banana diagram. This demonstrates how our general method is applied on explicit Feynman integrals and moreover shows the power of our method. For the reader the difficulty of our examples increases with the loop order and new appearing issues are highlighted and discussed case by case.

\subsection{Example 1: The Bubble Graph}
\label{1looptex} 

As the first example we discuss the one-loop banana diagram which is also called the bubble graph. This Feynman diagram can also be calculated directly with usual Feynman graph techniques\cite{Itzykson:1980rh}. Nevertheless, we will use for pedagogical reasons the bubble graph to introduce our method.

In our conventions the bubble integral is defined as
\begin{gleichung}
	\mathcal F_{\sigma_1}(t, \xi_1,\xi_2)	&=	\int_{x,y\geq 0} \frac{x \mathrm dy - y\mathrm dx}{xy\left(	t-(\xi_1^2x+\xi_2y^2)(\tfrac{1}{x} + \tfrac{1}{y})	\right)}	\\
							&=	-\xi_1\xi_2 \int_{x,y\geq 0} \frac{x \mathrm dy - y\mathrm dx}{x^2+uxy +y^2}~,
\label{bubble}
\end{gleichung}
where in the second line the coordinates are rescaled and $u = \tfrac{\xi_1^2+\xi_2^2-t}{\xi_1\xi_2}$ is introduced.

\begin{figure}[h!]
\begin{center}
\vspace{0.25cm}
\includegraphics[width=0.2\textwidth]{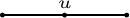}
\vspace{0.15cm}
\caption{The toric diagram for the bubble graph} 
 \label{toricl=1}
\end{center}
 \end{figure}

Following our method we associate to the bubble graph \eqref{bubble} the polynomial constrain
\begin{gleichung}
	P_1	=	x^2 + uxy + y^2
\label{polybubble}
\end{gleichung}
in projective space $\mathbb P$. For generic values of the parameter $u$ this defines two different points in $\mathbb P$. It looks a bit artificial but we can give a toric description of this algebraic variety consisting of two points. We take the Newton polytope of \eqref{polybubble} which is shown in Figure \ref{toricl=1}.
 
\noindent
It has a single $l$-vector and Batyrev coordinate
\begin{gleichung}
	l=(-2; 1,1) \quad\text{and}\quad	z= \frac{1}{u^2}~.
\label{batybubble}
\end{gleichung} 

As explained in section \ref{chapgeneral} we expect two functions spanning the function space of the bubble graph. One is coming from the maximal cut integral and the other one is a special solution of the inhomogeneous differential equation corresponding to the bubble graph. Furthermore, there is only a single true parameter for which we take naturally the Batyrev coordinate $z$ from \eqref{batybubble}.

The holomorphic period can be computed directly from the integral or from the $l$-vector \eqref{batybubble}
\begin{gleichung}
	\varpi	&=	\frac{1}{2 \pi i } \int_{S^1} \frac{x \mathrm dy - y\mathrm dx}{\sqrt{z}(x^2+y^2) + xy}	=	-\frac{1}{2 \pi i } \int_{S^1} \frac{1}{1 + \sqrt{z}(v+\tfrac1v)} \frac{\mathrm dv}{v}	\\
			&=	-\frac{1}{2 \pi i } \int_{S^1}\sum_{n=0}^\infty\sum_{m=0}^n (-1)^n\binom nm z^{n/2}v^{2m-n}\frac{\mathrm dv}{v}	=	-\sum_{n=0}^\infty	\frac{(2n)!}{(n!)^2}z^n	=	-\frac{1}{\sqrt{1-4z}}~,
\label{holbubble}
\end{gleichung}
where we have introduced the variable $v=\tfrac xy$. Moreover, $\varpi$ satisfies the first order differential equation
\begin{gleichung}
	\mathcal D\varpi	=	(1-4z)\theta\varpi	-  2z\varpi	=	0
\label{opbubble}
\end{gleichung}
with the logarithmic derivative $\theta = z\partial_z$.

Now we apply the operator $\mathcal D$ from \eqref{opbubble} on the integrand of the geometrical chain integral \eqref{opengeometric} containing the inner point of the polytope $u$ expressed through the Batyrev coordinate $z$. At the end we relate this expression to the bubble graph simply by dividing through the inner point. Fortunately, the integral in the bubble case can be computed analytically
\begin{gleichung}
	\mathcal D\Pi_{\sigma_1}	=	\mathcal D \int_{x,y\geq 0}	 u\frac{x \mathrm dy - y\mathrm dx}{x^2 + uxy + y^2}	&=	 \int_{x,y\geq 0} \mathcal D\frac{x \mathrm dy - y\mathrm dx}{\sqrt{z}(x^2+y^2) + x y}	=	1~.
\label{compinhom}
\end{gleichung}
This extends the homogeneous differential equation \eqref{opbubble} to an inhomogeneous one
\begin{gleichung}
	(1-4z)\theta\ \Pi_{\sigma_1}(z)	-  2z\ \Pi_{\sigma_1}(z)	=	1~.
\label{inhombubble}
\end{gleichung}
A special solution to this inhomogeneous differential equation is given by
\begin{gleichung}
	\varpi_S	=	\varpi\log(z)	+2z	+7z^2 + \tfrac{74}{3}z^3	+ \tfrac{533}{3}z^4 	+ \cdots	~.
\label{special}
\end{gleichung}

Then the general solution to the inhomogeneous differential equation \eqref{inhombubble} is given by $\Pi_{\sigma_1} = \varpi_S + \lambda \varpi$ with $\lambda \in \mathbb C$. We can relate this solution to the bubble graph by dividing with the inner point $u$ and rescaling it by $-\xi_1\xi_2$. The parameter $\lambda$ can be fixed by calculating the bubble graph \eqref{bubble} at a special point in moduli space, for example $u=1$. 

In the literature \cite{mixedhodgestructure} the $l$-loop banana diagrams were analyzed in the equal mass case, i.e. $\xi_i=1$ for $i=1, \hdots l+1$. The one-loop bubble diagram satisfies the inhomogeneous first order equation
\begin{gleichung}
	t(t-4)	f_1^\prime(t)	+(t-2) f_1(t)	=	-2!~.
\label{inhomhove}
\end{gleichung}
After dividing $\Pi_{\sigma_1}$ by the inner point this is exactly the differential equation it satisfies.

\subsection{Example 2: The Sunset Graph}
\label{sunsetex}

Our second example deals with the two-loop Banana diagram also known as the sunset graph. A different discussion of the sunset graph is given in \cite{MR3780269} from which we adopt parts of our notation. 

The sunset Feynman graph is defined by
\begin{gleichung}
	\mathcal F_{\sigma_2}(t, \xi_1,\xi_2,\xi_3) = \int_{\sigma_2} \frac{\mu_2}{P_2(t, \xi_1,\xi_2,\xi_3;x)}	=	\int_{\sigma_2} \frac{x \mathrm dy\wedge\mathrm dz-y \mathrm dx\wedge\mathrm dz+z \mathrm dx\wedge\mathrm dy}{xyz\left(t- (\xi_1^2x + \xi_2^2 y+\xi_3^2 z)(\tfrac1x + \tfrac1y + \tfrac1z)	\right)}~,
\label{sunset}
\end{gleichung}
with the integration domain defined in \eqref{sigmal}. It can be interpreted as a relative period on an elliptic curve defined by the polynomial constraint
\begin{gleichung}
	P_2	=	t x y z	-\xi _1^2 x^2 y-\xi _1^2 x^2 z-\xi _1^2 x y z	-\xi _2^2 x y^2-\xi_2^2 x y z-\xi _2^2 y^2 z		-\xi _3^2 x y z-\xi _3^2 x z^2-\xi _3^2 y z^2
\label{polyphysical}
\end{gleichung}
in an ambient space given by two-dimensional projective space $\mathbb P^2$ as explained in section \ref{ssec:geometry}. Our approach is strongly based on this geometric interpretation. For convenience we rescale the coordinates and introduce a simpler parametrization of the elliptic curve. The polynomial is then given as
\begin{gleichung}
	P_2	=	x y^2+y z^2+x^2z	+m_1x z^2+m_2x^2y+m_3y^2z		+u x y z~.
\label{polyum}
\end{gleichung}

\begin{figure}[h]
\begin{center}
\includegraphics[width=0.25\textwidth]{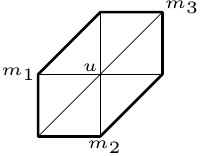}
 \caption{Toric diagram for the sunset graph}
     \label{toricl=2}
\end{center}
\end{figure}

We notice that the polynomial $\eqref{polyum}$ describes the blow up of $\mathbb P^2$ in three points which we call in the following $\mathcal E_{B3}$. In \cite{maximilianalbrecht} a nice analysis of the different blow ups of $\mathbb P^2$ is carried out from which we can extract same information for the toric description. In Figure \ref{toricl=2} the polyhedron corresponding to \eqref{polyum} is shown. The polyhedron's vertices are given by
\begin{gleichung}
	\nu_2 = \{ (0,-1) , (1,0) , (1,1) , (0,1) , (-1,0) , (-1,-1) , (0, 0) \}~.
\end{gleichung}
The corresponding Mori cone generators are given by
\begin{gleichung}
	\tilde l_1	&=	(1,-1,1,0,0,0, -1)~, \quad		&&\tilde l_2	=	(0,1,-1,1,0,0, -1) \\
	\tilde l_3	&=	(0,0,1,-1,1,0, -1)~, \quad		&&\tilde l_4	=	(0,0,0,1,-1,1, -1) \\
	\tilde l_5	&=	(1,0,0,0,1,-1, -1)~, \quad		&&\tilde l_6	=	(-1,1,0,0,0,1, -1)
\label{lnonsimplicial}
\end{gleichung}
generating a non-simplicial cone. From the general discussion in section \ref{ssec:geomperiods} we only need four independent $l$-vectors and also that in all columns of \eqref{lnonsimplicial} except the one corresponding to the inner point there is at least one positive entry. This does still not yield a distinct choice of four $l$-vectors but all of them can be used. We take for our collection of four $l$-vectors the ones which restrict to the Mori cone generators of the cubic in $\mathbb P^2$ and the other blow ups of $\mathbb P^2$ in one and two points. So we take in the following the four $l$-vectors
\begin{gleichung}
l_1=\tilde l_1~, \quad l_2=\tilde l_2~, \quad l_3=\tilde l_3 \quad\text{and}\quad l_4=\tilde l_4~.
\label{lsunset}
\end{gleichung}


Toric geometry singles out a natural choice of parametrization of the algebraic variety given by the Batyrev coordinates \eqref{zcoords}. These parameters are related to the ones in \eqref{polyphysical} and \eqref{polyum} by
\begin{gleichung}
	z_1	&=	-\tfrac{m_2m_3}{u} 	&=	-\tfrac{\xi _1^2}{\xi _1^2+\xi _2^2+\xi _3^2-t}~, \quad	z_2	&=	-\tfrac{1}{um_3}	&=	&-\tfrac{\xi _3^2}{\xi _1^2+\xi _2^2+\xi _3^2-t} \\
	z_3	&=	-\tfrac{m_1m_3}{u}	&=	-\tfrac{\xi _2^2}{\xi _1^2+\xi _2^2+\xi _3^2-t}~, \quad	z_4	&=	-\tfrac{1}{um_1}	&=	&-\tfrac{\xi _1^2}{\xi _1^2+\xi _2^2+\xi _3^2-t}~.
\label{batyrevsunset}
\end{gleichung}

Upon collecting the main toric information of our problem we can start with our strategy. The first part of our strategy will be the computation of the periods corresponding to the maximal cut integral.

Let us note here in passing that in the elliptic curve case it is not necessary to solve any differential equation to obtain the 
period integrals and hence the mass dependence of the ma\-xi\-mal cut integral. The periods are completely determined by 
modular functions as follows from~\cite{maximilianalbrecht}: We can bring  the constraint $P_2=0$ (\ref{polyum}) defining the 
elliptic curve into Weierstrass form   $y^2=4 x^3- x g_2(u,\underline{m})- g_3(u,\underline{m})$. This defines the modular 
parameter $\tau( u,\underline{m})$ from  the definition of the Hauptmodul $j$  of PSL$(2,\mathbb{Z})$   as 
\begin{gleichung}
	\frac{1728 g^2_3(u,\underline{m})}{g^3_2(u,\underline{m})-27 g_3^2(u,\underline{m})}	=j=	\frac{1}{q} + 744+ \num{192688} q + 21493760q^2+{\cal O}(q^3)~,
\end{gleichung}
where $q=\exp( 2\pi i \tau)$.   Then the period $\int_a \Omega/u$  which yields  the  
maximal cut integral is given in terms of the Eisenstein series as  
\begin{gleichung}
	\partial_u t(u)	=	\int_a \Omega=\sqrt{\frac{E_6(\tau(u,\underline{m})) g_2(u,\underline{m})}{E_4(\tau(u,\underline{m})) g_3(u,\underline{m})}}~.
\end{gleichung}
Moreover, the dual period $\int_b \Omega$ can be obtained by special geometry  of non-compact three-folds  as  
$\partial_t \int_b\Omega=-\frac{1}{2 \pi i} \tau( u,\underline{m}))=\partial_t^2 F(t)$, where $F$ is 
the prepotential that features in local mirror symmetry as generating function for the genus zero BPS  invariants $n_0^\beta$, which is given by
\begin{equation} 
F(Q)=-\frac{c^{ijk}}{3!} t_i t_j t_k + \frac{c^{ij}}{2} t_i t_j +c^i t_i + c \sum_{\beta\in H_2(W,\mathbb{Z})} n_0^\beta {\rm Li}_3(Q^\beta)\ .
\end{equation} 
Here $t_i$ are the flat coordinates, $Q_i=\exp(t_i/2\pi i)$ and the  $c^*$ are classical intersection numbers  on the mirror $W$. 
In~\cite{maximilianalbrecht} the K\"ahler classes $t_i$ for $i=1,\ldots,4$   of the mirror have been identified. These  are linearly 
related to the Batyrev coordinates (\ref{batyrevsunset}). With $Q_i =\exp(t_i/2 \pi i)$ they relate to the physical parameters as   
\begin{equation} 
Q=(Q_1 Q_2 Q_3 Q_4)^\frac{1}{3}~, \quad m_1=\frac{(Q_1 Q_3 Q_4)^\frac{1}{3}}{Q_2^\frac{2}{3}}~, \quad m_2=\frac{(Q_1 Q_2 Q_4)^\frac{1}{3}}{Q_3^\frac{2}{3}}~, \quad m_3=\frac{(Q_1 Q_2 Q_3)^\frac{1}{3}}{Q_4^\frac{2}{3}}\ . 
\label{Kahler}  
\end{equation} 
This allows to relate the full integer genus zero BPS expansion $n^\beta_0$ in the four  K\"ahler 
pa\-ra\-me\-ters~\cite{maximilianalbrecht}
\begin{equation}
\begin{array}{rl}
 F&	=  cl. + L_{0,0,0,1} + L_{1,0,0,1} - 2L_{1,0,1,1} + 3L_{1,1,1,1} + 3L_{2,1,1,1} - 4L_{2,1,1,2} + 5L_{2,1,2,2} \\ &
 	 -6L_{2,2,2,2} + 5L_{3,1,2,2} - 6L_{3,1,2,3} + 7L_{3,1,3,3} - 36L_{3,2,2,2} + 35L_{3,2,2,3} - 32L_{3,2,3,3} \\ &
	 +27L_{3,3,3,3} + 7L_{4,1,3,3} - 8L_{4,1,3,4} + 9L_{4,1,4,4} - 6L_{4,2,2,2} + 35L_{4,2,2,3} \\ &
	 - 32L_{4,2,2,4} -160L_{4,2,3,3} + 135L_{4,2,3,4} - 110L_{4,2,4,4}+ 531L_{4,3,3,3} \\&  
	 -400L_{4,3,3,4} + 286L_{4,3,4,4} -192L_{4,4,4,4} +{\rm Sym}_{ijk}(L_{a,i,j,k})+\cdots  
 \end{array}
\end{equation} 
to the full set of physical parameters.  Here $L_\beta\coloneqq {\rm Li}_3(\prod_{i=1}^4 Q_i^{\beta_i})$. In~\cite{Bloch:2016izu}  BPS invariants are 
given for the projective parametrization $n_{ijk}$. The relation to the geometrical BPS invariants is 
$\sum_a  n_0^{aijk}=n_{ijk}$. It is clear from (\ref{Kahler})  and the symmetries of the polytop that the last formula 
is symmetric in the $ijk$ indices. Moreover, the  one parameter specialization also noted in~\cite{Bloch:2016izu} is given by $n_d= \sum_{a,i+j+k=d} n_0^{aijk}$. 
While we think that in the elliptic two-loop case  this relation of the  BPS expansion  to the Feynman graph is remarkable 
but not very useful, it becomes more useful for the higher loop banana graphs as we explain in section \ref{equalmassK3}.

\subsubsection{The sunset maximal cut integral}
The maximal cut integral of the sunset graph $\mathcal F_{T^2}(t, \xi_1,\xi_2,\xi_3)$ is defined by replacing the simplex $\sigma_2$ by a torus $T^2$. Instead of focusing on the maximal cut Feynman graph we rather deal with the related geometrical period which includes additionally the inner point $u$ of the toric diagram. The expression
\begin{gleichung}
	\Pi(T^2)(u, m_1,m_2,m_3) = \int_{T^2} u\frac{x \mathrm dy\wedge\mathrm dz-y \mathrm dx\wedge\mathrm dz+z \mathrm dx\wedge\mathrm dy}{	x y^2+y z^2+x^2z	+m_1x z^2+m_2x^2y+m_3y^2z		+u x y z	}
\label{maxcutsunset}
\end{gleichung}
describes a ``usual'' period on the elliptic curve $\mathcal E_{B3}$ and it is easily related to the maximal cut integral $\mathcal F_{T^2}$ by dividing with $u$. At the point of maximal unipotent monodromy the geometrical period $\Pi(T^2)=\varpi$ is given by a single holomorphic power series \eqref{varpi}. Evaluating the period \eqref{maxcutsunset} at a generic point in moduli space requires the knowledge of a period basis. Such a period basis can be found as follows: Homology theory of a generic elliptic curve tells us that there exists only a pair of one-cycles, i.e. $H_2(T^2) = \mathbb Z^2$. So if we take the $(1,0)$-form $\frac{a_0\mu_2}{P_2}$ with $a_0$ the inner point of the polytope and $P_2$ the hypersurface constraint defining the elliptic curve there are only two independent periods. Here it is important to remark that for elliptic curves this statement is independent of the parametrization, in particular, independent of the number of moduli. For the geometrical period $\Pi(T^2)$ and therefore also for the maximal cut integral $\mathcal F_{T^2}$ this means that there are two independent functions which linearly combined yield \eqref{maxcutsunset} at a generic point in moduli space.

In our toric analysis it is convenient to use the Batyrev parameters defined in \eqref{batyrevsunset}. Later we will see that the usage of this particular choice of parametrization enables us to fully determine the sunset graph. Moreover, it simplifies many of the subsequent results.

From the Mori cone generators \eqref{lsunset} one can directly write down the holomorphic period at the point of maximal unipotent monodromy given by
\begin{gleichung}
	\varpi(\underline z)	&=	\sum_{\underline m\geq0} \frac{ \Gamma
   \left(1+ m_1+m_2+m_3+m_4\right)}{\Gamma \left(1+m_1\right) \Gamma
   \left(1-m_1+m_2\right) \Gamma \left(1+m_1-m_2+m_3\right) \Gamma
   \left(1+m_3-m_4\right)} \\
   						&\quad \cdot \frac{1}{ \Gamma \left(1+m_4\right) \Gamma \left(1+m_2-m_3+m_4\right)}z_1^{m_1} z_2^{m_2} z_3^{m_3} z_4^{m_4}
\label{foursunset}
\end{gleichung}
with the abbreviations $\underline z= (z_1,z_2,z_3,z_4)$ and $\underline m= (m_1,m_2,m_3,m_4)$. This is the most generic four-parameter holomorphic period of $\mathcal E_{B3}$. The geometrical period \eqref{maxcutsunset} has one less parameter since one-parameter can be scaled away. Therefore, we have to specialize the four-parameter solution \eqref{foursunset} to a three-parameter one. We remark that from \eqref{batyrevsunset} the parameters $z_1$ and $z_4$ have the same value if expressed in the physical parameters. This means that the four-parameter solution \eqref{foursunset} specialized on the subslice with $z_1=z_4$ corresponds to the holomorphic solution of the geometrical period \eqref{maxcutsunset} at the maximal unipotent monodromy point.

This subslice is not as problematic as for the higher loop banana graphs because the sum over $m_4$ still contains a parameter, here $z_1$. But still we can use the $\Gamma$-functions in \eqref{foursunset} to bound the summation over the index $m_4$ by $m_3-m_4 \geq 0$. We obtain for the first few orders
\begin{gleichung}
	\varpi(z_1,z_2,z_3)	&=	1 + 2z_1z_2+2z_1z_3+2z_2z_3	+	12 z_1z_2z_3	\\
						&\quad	+	6 z_1^2z_2^2 +24z_1^2z_2z_3+24z_1z_2^2z_3+6z_1^2z_3^2+24z_1z_2z_3^2 + 6z_2^2z_3^2 + \cdots~.
\label{regsolutionsunset}
\end{gleichung}

Now our strategy is as follows: We compute the holomorphic solution to high order such that we can find a set of differential operators annihilating it. This set of differential operators has to be complete in a sense that its solutions form a basis of period integrals on the elliptic curve $\mathcal E_{B3}$. Therefore, a suitable ansatz for these operators is crucial. Again homology theory of the elliptic curve tells us what kind of solutions we expect and so the rare form of the operators. For $\mathcal E_{B3}$ only two solutions exist. At the point of maximal unipotent monodromy the analytic structure of them is also known. One is a holomorphic function in the parameters and the other contains single logarithms of the parameters. For the differential operator ideal this implies that we are searching for first order operators in the parameters $(z_1,z_2,z_3)$. Having found the first few operators one has to increase the number of operators until they are enough to fully determine the two different periods. As a possible generating set of the ideal we find
\begin{gleichung}
	\mathcal D_1	&=	\theta _1-\theta _2+z_2 \left(\theta _1-\theta _2+\theta _3+2 z_3\left(\theta _1+\theta
   _2+\theta _3+1\right) \right)	\\
   				&\quad -z_1 \left(-\theta _1+\theta _2+\theta _3+2z_3
   \left(\theta _1+\theta _2+\theta _3+1\right) \right)		\\
   	\mathcal D_2	&=	\theta _2-\theta _3+z_3\left(\theta _1+\theta _2-\theta _3\right) -z_2\left(\theta
   _1-\theta _2+\theta _3\right) -2 z_1
   \left(z_2-z_3\right)
\left(\theta _1+\theta _2+\theta _3+1\right) 		\\
	\mathcal D_3	&=	\left(\theta _1-\theta _2\right) \left(\theta _1+\theta _2-\theta _3\right)+z_1\left(\theta
   _1-\theta _2-\theta _3\right) \left(\theta _1+\theta _2+\theta _3+1\right)	\\
   				&\quad   +z_2\left(\theta _1-\theta _2+\theta _3\right) \left(\theta _1+\theta _2+\theta
   _3+1\right) 
\label{opsunset}
\end{gleichung}
with $\theta_i = z_i \partial_{z_i}$ for $i=1,2,3$. The missing period is then given by
\begin{equation}
	\Pi(\Gamma_1)(z_1,z_2,z_3)	= \varpi\left(\log(z_1) + \log(z_2) + \log(z_3)\right) + \Sigma_1	
\label{logsolutionsunset}
\end{equation}
with
\begin{gleichung}
	\Sigma_1		&=	z_1 + z_2 + z_3 -\tfrac{z_1^2}{2}+7 z_1 z_2+7 z_1 z_3-\tfrac{z_2^2}{2}+7 z_2 z_3-\tfrac{z_3^2}{2}  \\
				&\quad +\tfrac{z_1^3}{3}+3 z_1^2z_2 +3  z_1^2z_3+3 z_1z_2^2 +3 z_1z_3^2 +48 z_1z_2 z_3
   +\tfrac{z_2^3}{3}+\tfrac{z_3^3}{3}+3 z_2 z_3^2+3 z_2^2 z_3 + \cdots~.
\end{gleichung}
These two solutions \eqref{regsolutionsunset} and \eqref{logsolutionsunset} form a basis of the periods for the elliptic curve $\mathcal E_{B3}$. Using the relations \eqref{batyrevsunset} we can divide by the inner point and transform this basis to the necessary point in moduli space such that they can be linearly combined to yield the maximal cut integral $\mathcal F_{T^2}$.

In the next section we extend the differential operator ideal \eqref{opsunset} such that it governs all functions describing the full geometrical sunset Feynman graph $\Pi_{\sigma_2}$. By dividing with the inner point we can transfer these results to the actual Feynman integral \eqref{sunset}.

\subsubsection{Extension to inhomogeneous differential operators}

As explained in section \ref{inhomstrategy} we find as the first step the inhomogeneities of the operators \eqref{opsunset}. Again we use the Batyrev coordinates $(z_1,z_2,z_3)$ which is crucial for the applicability of our method. We apply the operators \eqref{opsunset} on the geometrical differential $\tfrac{u\mu_2}{P_2}$ and integrate afterwards over the two-dimensional simplex $\sigma_2$. These chain integrals can not in general be computed analytically with generic parameters but numerical evaluations of these integrals for fixed values of the parameters are possible. Now the advantage of the Batyrev coordinates is that we can guess the exact values of the numerical results. We claim that the differential operator ideal only produces simple logarithmic expressions in the Batyrev coordinates $(z_1,z_2,z_3)$. For \eqref{opsunset} we find the following inhomogeneities\footnote{We checked this numerically up to more than 15 digits and for different values of the variables $z_i$ for $i=1,2,3$.}
\begin{gleichung}
	\mathcal D_1 \Pi_{\sigma_2}	&=	- \log(z_2) + \log(z_3) \\
	\mathcal D_2 \Pi_{\sigma_2}	&=	- \log(z_1) + \log(z_2) \\
	\mathcal D_3 \Pi_{\sigma_2}	&=	0 ~.
\label{inhomsunset}
\end{gleichung}

We think that in another parametrization, for instance the physical parameters $(t, \xi_1,\xi_2,\xi_3)$, and without the inner point these integrals can neither be computed analytically nor their numerical values can be guessed. Only the geometrical differential in the special parametrization with the Batyrev parameters guarantees the feasibility of our method.

Having found the complete set of inhomogeneous differential operators their solutions can be computed easily. One has to extend the solutions of the homogeneous system \eqref{opsunset} by a special solution satisfying \eqref{inhomsunset}. As an ansatz for this solution we increase the power of logarithms in $(z_1,z_2,z_3)$ up to two. Then we find as a possible choice of special solution
\begin{gleichung}
	\varpi_S(z_1,z_2,z_3)	&=	\left(	\log(z_1)\log(z_2)+\log(z_1)\log(z_3)+\log(z_2)\log(z_3)	\right) \varpi_0	\\
						&\hspace{-1.5cm} +2\log(z_1)+2\log(z_2)+2\log(z_3)		+	2z_1\log(z_1)+2z_2\log(z_2)+2z_3\log(z_3)	\\
						&\hspace{-1.5cm}	-\tfrac{z_1^2}{2}+10z_1z_2-\tfrac{z_2^2}{2}+10z_1z_3-\tfrac{z_3^2}{2}-z_1^2\log(z_1)+10z_1z_2\log(z_1) \\
						&\hspace{-1.5cm} +10z_1z_3\log(z_1)+6z_2z_3\log(z_1) +10z_1z_2\log(z_2)-z_2^2\log(z_2)+6z_1z_3\log(z_2)	\\
						&\hspace{-1.5cm} +10z_2z_3\log(z_2) +6z_1z_3\log(z_3)+10z_1z_3\log(z_3)+10z_2z_3\log(z_3)-z_3^2\log(z_3) +\cdots~.
\label{specialsolutionsunset}
\end{gleichung}
\indent
The general solution is then a linear combination of the form $\Pi_{\sigma_2} = \varpi_S + \lambda_0 \varpi + \lambda_1 \Pi(\Gamma_1)$ with $\lambda_0, \lambda_1 \in \mathbb C$. We can express $\Pi_{\sigma_2}$ through the physical parameters $(t, \xi_1,\xi_2,\xi_3)$ and divide it by the inner point to find the full sunset Feynman graph $\mathcal F_{\sigma_2}$ \eqref{sunset}.

\subsubsection{Comparison with the equal mass case and other known results}
\label{subsecresultssunset}
Many results about the sunset graph are already known in the literature \cite{Bogner:2010kv,Broedel:2019kmn}. In particular, the equal mass case 
meaning $\xi_i = 1~\text{for } i=1,2,3$ was analyzed many times. In this case, the maximal cut integral is up to a factor of $u=t-3$ (\ref{coefficientinnerpoint})  
the holomorphic  period of the Barth-Nieto elliptic curve that can be represented as in \eqref{BarthNieto}.  The equal mass sunset graph has 
to satisfy an inhomogeneous second order differential equation \cite{mixedhodgestructure} in the momentum variable $t$
\begin{gleichung}
	t(t-1)(t-9)f''_2(t)+(3t^2-20t+9)f'_2(t)+(t-3)f_2(t)	=	-3!~.
\label{opequalsunset}
\end{gleichung}
Our three-parameter solutions \eqref{regsolutionsunset}, \eqref{logsolutionsunset} and \eqref{specialsolutionsunset} break down in the equal mass case\footnote{Notice that before one can apply the differential equation \eqref{opequalsunset} on our solutions they have to be transformed at the same point in moduli space, which is here $t \mapsto \frac 1t$.} to the solutions of \eqref{opequalsunset}. This shows that they reproduces the well established equal mass results.

For the sunset graph a second test is possible since in \cite{MR3780269} an inhomogeneous differential equation in all physical parameters is given which the sunset graph has to satisfy. Here we notice that our holomorphic and single logarithmic solutions expressed in the physical parameters fulfill this equation. The special solution \eqref{specialsolutionsunset} does not. A direct comparison between our special solution and the solutions to the inhomogeneous differential equation in \cite{MR3780269} shows that the discrepancy between them is only in the terms having no logarithm in the variable $s=1/t$. Such a small difference can be a result of a typo in the polynomials given in \cite{MR3780269} but a general mistake in their derivation of the inhomogeneous differential equation can not be excluded.

\begin{table}[h]
\centering
\begin{tabular}{@{}c | cc c@{}}
\toprule
	$\Pi_{\sigma_2} = \lambda_S\varpi_S+\lambda_0\varpi_0+\lambda_1\Pi(\Gamma_1)$		& $\lambda_S$	 & $\lambda_0$  & $\lambda_1$ 	\\ \midrule
   order 5 		& $\num{0.9998}$	& $\num{-29.6275+42.7536 i}$	& $\num{-13.6122-18.8466i}$				\\
   order 10   	& $\num{1.0000}$	& $\num{-29.6088+42.7407 i}$	& $\num{-13.6048-18.8496i}$   				\\ \midrule
   order 5		& $\num{1.0004+0.0007i}$	& $\num{70.0913+109.3340i}$	& $\num{-34.7859-18.8389i}$		\\
   order 10		& $\num{1.0004+0.0007i}$	& $\num{70.0913+109.3340i}$	& $\num{-34.7859-18.8389i}$		\\
\bottomrule
\end{tabular}
\caption{Linear combination of solutions for the sunset graph. In the first two rows are the values for our solutions whereas the last two give the ones for the solutions from \cite{MR3780269}.}
\label{tab1}
\end{table}

To demonstrate the correctness of our solutions we made some numerical checks. We evaluated the sunset Feynman graph \eqref{sunset} at three different points\footnote{We took for the three points the values $(s,\xi_1,\xi_2,\xi_3)=(s_1+i/10,1/10,1/20,1/30)$, for $s_1=1/10,~s_2=1/20$ and $s_3=1/30$.} to fix the linear combination of our three solutions\footnote{We fixed our basis of solutions such that the holomorphic solution starts with one and the constant piece in the single logarithmic solution is zero. Moreover, we fixed the special solution by requiring that the constant term and the constant term multiplied by $\log s$ is vanishing.}. Having found the right combination of solutions given in Table \ref{tab1} we checked for further values of the parameters and compare the precision for different expansion orders of $\varpi, \Pi(\Gamma_1)$ and $\varpi_S$. Our results are listed in Table \ref{tab2}. Notice, that it is important that the value of one $\xi_i$ is fixed since there are only three physical degrees of freedom after rescaling. We choose $\xi_3$ to be fixed. With increasing expansion order our solutions fit better and better to the sunset graph which we could not observe for the solutions of \cite{MR3780269}. Moreover, the factor $\lambda_S$ of the special solution $\varpi_S$ tends to the value one as expected.

\begin{table}[h]
\centering
\begin{tabular}{@{}c | cc | cc@{}}
\toprule
$s,\xi_1,\xi_2,\xi_3,$ & order 5 & order 10  & order 5 & order 10 \\ \midrule
   1/27 + i/20,   1/10, 1/20, 1/30    &  $9\cdot10^{-5}$      &    $5\cdot10^{-9}$      &   $2\cdot10^{-4}$      &   $2\cdot10^{-4}$   	\\
    1/21+i/10, 1/10, 1/50, 1/30      &    $4\cdot10^{-4}$     &     $6\cdot10^{-9}$     &    30   			  &     30   				\\
    1/24+i/10, 1/10+i/15,1/20,1/30	& $6\cdot10^{-4}$ 	& $5\cdot10^{-9}$	&    22			  &	22				\\
\bottomrule
\end{tabular}
\caption{The table shows how precise the relative periods combined as listed in Table \ref{tab1} describe the Feynman graph. We show the absolute value of the difference between the numerical computation of the sunset graph and the evaluation of the linear combination of solutions. Increasing the expansion order increases the precision of our results given as the second and third column. The last columns give the results from \cite{MR3780269} which do not increase their precision.}
\label{tab2}
\end{table}

\subsection{Example 3: The Three-Loop Banana Graph}
\label{3looptex}
As our last and most complicated example we demonstrate the applicability of our approach for the three-loop banana diagram
\begin{gleichung}
	\mathcal F_{\sigma_3}(t, \xi_1,\xi_2,\xi_3,\xi_4) = \int_{\sigma_3}\frac{x \mathrm dy\wedge\mathrm dz\wedge \mathrm dw-y \mathrm dx\wedge\mathrm dz\wedge \mathrm dw+z \mathrm dx\wedge\mathrm dy\wedge \mathrm dw - w \mathrm dx\wedge \mathrm dy \wedge \mathrm dz}{xyzw\left(t- (\xi_1^2x + \xi_2^2 y+\xi_3^2 z+\xi_4^2 w)(\frac1x + \frac1y + \frac1z+\frac1w)	\right)}~.
\label{banana}
\end{gleichung}
The three-loop banana Feynman graph \eqref{banana} can again be interpreted as a relative period now on a $\mathrm K3$ surface. This $\mathrm K3$ surface is defined by the constraint $P_3$ from the denominator in \eqref{banana}. After a rescaling of the coordinates we obtain
\begin{gleichung}
	P_3	&=	x^2yz + xyw^2 + xzw^2 + yzw^2 + m_1xy^2w + m_2x^2zw + m_3yz^2w + m_4x^2yw \\
				&\quad + m_5xz^2w + m_6y^2zw + m_7xy^2z + m_8xyz^2 + uxyzw~.
\label{poly3um}
\end{gleichung}

\begin{figure}[h!]
\begin{center}
\includegraphics[width=0.5\textwidth]{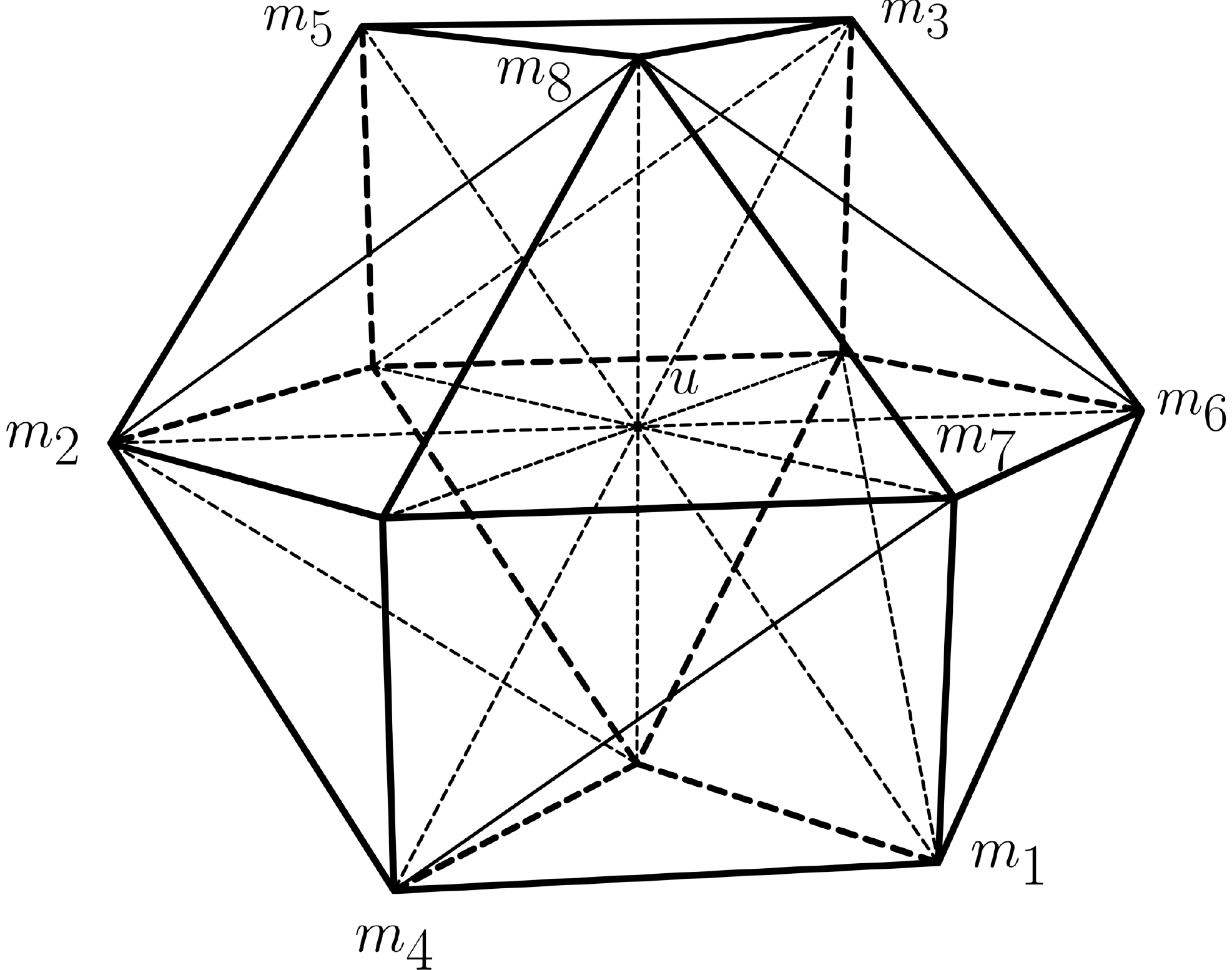}
 \caption{Toric diagram for the three-loop banana graph}
     \label{toricl=3}
\end{center}
\end{figure}

The polytope $P_{\Delta_3}$ corresponding to the banana graph together with a triangulation is shown in Figure \ref{toricl=3}. Its vertices are given by
\begin{gleichung}
	\nu_3	=	&\{	(-1, 1, 0), (1, 0, 0), (0, -1, 1), (0, 0, 1), (1, -1, 0), (1, 0, -1), (0, 0, -1), (-1, 0, 1),  \\
			&\quad	(0, -1, 0), (-1, 0, 0), (0, 1, 0), (0, 1, -1), (0, 0, 0)	\}~.
\end{gleichung}
Furthermore, the Mori cone generators corresponding to the triangulation drawn in the polytope in Figure \ref{toricl=3} are given by
\begin{gleichung}
	l_1	&=	(0, 0, -1, 0, 1, 0, 0, 1, 0, 0, 0, 0, -1)~,		\quad l_2	&&=	(0, -1, 0, 0, 1, 0, 0, 0, 0, 0, 1, 0, -1) \\
	l_3	&=	(0, -1, 0, 1, 0, 1, 0, 0, 0, 0, 0, 0, -1)~,		\quad l_4	&&=	(0, 0, -1, 1, 0, 0, 0, 0, 1, 0, 0, 0, -1) \\
	l_5	&=	(-1, 0, 0, -1, 0, 0, 0, 1, 0, 0, 1, 0, 0)~,		\quad l_6	&&=	(0, 0, 0, 0, -1, 1, -1, 0, 1, 0, 0, 0, 0) \\
	l_7	&=	(0, 0, 1, 0, 0, 0, 0, -1, -1, 1, 0, 0, 0)~,		\quad l_8	&&=	(0, 1, 0, 0, 0, -1, 0, 0, 0, 0, -1, 1, 0) \\
	l_9	&=	(1, 0, 0, 0, 0, 0, 1, 0, 0, -1, 0, -1, 0)~.									
\label{lk3}
\end{gleichung}
They form a simplicial Mori cone generated by $3^2$ vectors. For the subsequent discussion we need the Batyrev coordinates together with their relations to the physical paramters
\begin{gleichung}
 z_1 &= -\tfrac{m_2 m_3}{m_5 u} 	&= -\tfrac{\xi _1^2}{\xi _1^2+\xi _2^2+\xi _3^2+\xi _4^2-t}~, \qquad	 z_5 &= \tfrac{m_3 m_7}{m_6 m_8} 	&=&~ 1	\\
 z_2 &= -\tfrac{m_2 m_7}{u} 		&= -\tfrac{\xi _2^2}{\xi _1^2+\xi _2^2+\xi _3^2+\xi _4^2-t}~, \qquad	 z_6 &= \tfrac{m_4}{m_2}			&=&~ 1 \\
 z_3 &= -\tfrac{m_4 m_8}{u} 		&= -\tfrac{\xi _3^2}{\xi _1^2+\xi _2^2+\xi _3^2+\xi _4^2-t}~, \qquad	 z_7 &= \tfrac{m_5}{m_3} 			&=&~ 1 \\
 z_4 &= -\tfrac{m_8}{m_5 u} 		&= -\tfrac{\xi _4^2}{\xi _1^2+\xi _2^2+\xi _3^2+\xi _4^2-t}~, \qquad	 z_8 &= \tfrac{m_1}{m_4 m_7} 		&=&~ 1 \\
 \textcolor{white}{z_4} & \textcolor{white}{= -\tfrac{m_8}{m_5 u} }		& \textcolor{white}{= -\tfrac{\xi _4^2}{\xi _1^2+\xi _2^2+\xi _3^2+\xi _4^2-t}~, }\quad	 %
 																				z_9 &= \tfrac{m_6}{m_1} 			&=&~ 1~.
\label{batyrevk3}
\end{gleichung}

Having defined the most important information about the three-loop banana graph we want to find a set of functions describing it. We follow our general strategy but there are some subtleties which have not popped up for the sunset graph.

\subsubsection{Maximal cut integral}
As before, the maximal cut integral $\mathcal F_{T^3}(t, \xi_1,\xi_2,\xi_3,\xi_4)$ is related through the inner point to the $\mathrm K3$ period integral
\begin{gleichung}
	\Pi(T^3)(u, m_1,m_2,m_3,m_4) = \int_{T^3} \frac{u \mu_3}{P_3}~.
\label{maxcutk3}
\end{gleichung} 
We want to compute a basis for the periods on the $\mathrm K3$ surface. Cohomology theory of the $\mathrm K3$ surface can tell us again how many independent periods we expect. Differently as for elliptic curves the number of independent two-cycles depends on the number of moduli. For a $r$ parameter model we expect $r+2$ independent two-cycles and similarly $r+2$ independent periods. Moreover, the analytic structure of these periods can be specified further. There is exactly one holomorphic and one double logarithmic period on the $\mathrm K3$. The remaining $r$ periods are single logarithmic ones. 

The starting point of our method is the holomorphic period expressed through the Batyrev parameters which are much more as the physical parameters. From \eqref{batyrevk3} five Batyrev parameters are set to one after identification with the physical parameters. The remaining four coordinates $(z_1,z_2,z_3,z_4)$ are related to the physical parameters and are such the only ones important in the following. From the Mori cone generators it is always possible to write down the general form of the holomorphic period but in all nine Batyrev parameters. We can expand this holomorphic solution in the ``unphysical'' parameters $(z_5,z_6,z_7,z_8,z_9)$ exactly and set them afterwards to one. This yields the holomorphic solution in the physically relevant four parameters. To insure that our expansion is exact in the unphysical parameters we use the particular form of the holomorphic periods in terms of $\Gamma$-functions. Since the numerator does never diverge for positive values of the index parameters $m_i,~i=1,\hdots 9$ the $\Gamma$-functions in the denominator give bounds on the index parameters $m_i$. Concretely we obtain
\begin{gleichung}
	\varpi(z_1,z_2,z_3,z_4)	=	\sum_{\mathcal M} &\tfrac{ \Gamma
   \left(1+m_1+m_2+m_3+m_4\right)}{\Gamma \left(1+m_3+m_4-m_5\right) \Gamma
   \left(1+m_1+m_2-m_6\right) \Gamma \left(1+m_1+m_5-m_7\right) \Gamma
   \left(1+m_4+m_6-m_7\right)  } \\
   	&\cdot \tfrac{z_2^{m_2} z_3^{m_3} z_4^{m_4} z_5^{m_5} }{\Gamma \left(1-m_1-m_4+m_7\right) \Gamma
   \left(1+m_2+m_5-m_8\right) \Gamma \left(1+m_3+m_6-m_8\right) \Gamma
   \left(1-m_2-m_3+m_8\right)} \\
   	&\cdot \tfrac{1}{ \Gamma \left(1+m_7-m_9\right) \Gamma
   \left(1+m_8-m_9\right) \Gamma \left(1-m_5+m_9\right) \Gamma \left(1-m_6+m_9\right)}
\end{gleichung}
with the summation range given by
\begin{gleichung}
	\mathcal M	&=	\{	0\leq m_1 \leq \infty,\ 0\leq m_2 \leq \infty,\ 0\leq m_3 \leq \infty,\ 0\leq m_4 \leq \infty,\ 0\leq m_5 \leq m_3+m_4, \\
				&\quad	m_2+m_3\leq m_8 \leq m_2+m_5,\ 0\leq m_6 \leq m_1+m_2,\ m_1+ m_4 \leq m_7 \leq m_1+m_5, \\
				&\quad	m_6\leq m_9 \leq m_7 	\}~.
\end{gleichung}
We find
\begin{gleichung}
	\varpi(z_1,z_2,z_3,z_4)	&=	1	+	2 \left(	z_1 z_2+z_1 z_3+z_1 z_4	+	z_2 z_3+z_2 z_4	+	z_3 z_4	\right)	\\
						&\quad +	12 \left(	z_1 z_2 z_3+z_1 z_2 z_4+z_1 z_3 z_4+z_2 z_3 z_4		\right)		+	\cdots~.
\label{regsolutionk3}
\end{gleichung}
Then our strategy is the same as before. We expand the holomorphic solution \eqref{regsolutionk3} high enough that we can find a set of operators annihilating it. This time we are looking for second order operators in such a way that their solutions are given by a single holomorphic and a single double logarithmic solution and further four single logarithmic solutions. As a choice we take the operators $\mathcal D_1, \hdots, \mathcal D_4$ as generators for the differential operator ideal. They are listed in appendix \ref{appop}.
Then a period basis is given by four single logarithmic solutions
\begin{gleichung}
	\Pi(\Gamma_1^1)		&=	\varpi\log(z_1)	+\Sigma_1^1	\\
	\Pi(\Gamma_1^2)		&=	\varpi\log(z_2)	+\Sigma_1^2	\\
	\Pi(\Gamma_1^3)		&=	\varpi\log(z_3)	+\Sigma_1^3	\\
	\Pi(\Gamma_1^4)		&=	\varpi\log(z_4)	+\Sigma_1^4~,
\label{logsolutions}
\end{gleichung}
with
\begin{gleichung}
	\Sigma_1^1	&=	-z_1+z_2+z_3+z_4	+  \tfrac{z_1^2}{2}+z_1z_2 +z_1z_3 +z_1z_4 -\tfrac{z_2^2}{2} +z_2 z_3+5 z_2 z_4-\tfrac{z_3^2}{2}+5 z_3 z_4 -\tfrac{z_4^2}{2}	\\
				&\quad -\tfrac{z_1^3}{3}-3 z_1^2z_2 -3 z_1^2z_3 -3 z_1^2z_4 +3 z_1z_2^2 +3 z_1z_3^2 +3 z_1z_4^2 +16 z_1z_2 z_3 +16 z_1z_2 z_4 	\\
   				&\quad +16 z_1z_3 z_4 +\tfrac{z_2^3}{3}+3 z_2^2 z_3+3 z_2^2 z_4+3 z_2 z_3^2+3 z_2 z_4^2+52 z_2 z_3 z_4 +\tfrac{z_3^3}{3}+3 z_3^2 z_4\\
   				&\quad +3 z_3 z_4^2+\tfrac{z_4^3}{3}		+ \cdots~.
\end{gleichung}
The other $\Sigma_1^i$ for $i=2,3,4$ are given as permutations, namely  $\Sigma_1^2=\Sigma_1^1(z_1\leftrightarrow z_2)$, $\Sigma_1^3=\Sigma_1^1(z_1\leftrightarrow z_3)$ and $\Sigma_1^4=\Sigma_1^1(z_1\leftrightarrow z_4)$.
Additionally, there is a double logarithmic solution
\begin{gleichung}
	\Pi(\Gamma_2)		&=	\varpi \left[ \log(z_1)\log(z_2) + \log(z_1)\log(z_3) + \log(z_1)\log(z_4) + \log(z_2)\log(z_3)		\right.	\\
   					&\qquad	\left.	+ \log(z_2)\log(z_4) + \log(z_3)\log(z_4) 	\right]	+ \left(	\Sigma_1^2+\Sigma_1^3+\Sigma_1^4 \right)\log(z_1)	\\
					&\quad	+ \left(	\Sigma_1^1+\Sigma_1^3+\Sigma_1^4 \right) \log(z_2) + \left(	\Sigma_1^1+\Sigma_1^2+\Sigma_1^4 \right)\log(z_3)	\\
					&\quad	+ \left(	\Sigma_1^1+\Sigma_1^2+\Sigma_1^3 \right) \log(z_4)	+	\Sigma_2
\label{doublelogsolution}
\end{gleichung}
with
\begin{gleichung}
	\Sigma_2		&=	4 \left(z_1 z_2+z_3 z_2+z_4 z_2+z_1 z_3+z_1 z_4+z_3 z_4\right)	+ 6 \left( 2  z_1^2 z_2+2 z_1^2 z_3 +2 z_1^2 z_4  +2 z_1z_2^2 	\right.	\\
				&\quad \left.	+2 z_1 z_3^2 +2 z_1 z_4^2 +11 z_2    z_3 z_1+11 z_1 z_2 z_4 +11 z_1z_3 z_4 +2 z_2 z_3^2	\right.	\\
 				&\quad \left.	+2 z_2 z_4^2 +2 z_3 z_4^2 +2  z_2^2 z_3 +2 z_2^2 z_4+2 z_3^2 z_4+11 z_2 z_3 z_4		\right)		+ \cdots~.
\end{gleichung}
Together with the holomorphic period \eqref{regsolutionk3} this completes the period basis. 

There is another very compact way of expressing the double logarithmic solution. We define the so called mirror maps
\begin{gleichung}
	t_i	&=	\tfrac{\Pi(\Gamma_1^i)}{2\pi i\varpi}	\quad\text{for } i=1, \hdots, 4~.
\label{mirrormaps}
\end{gleichung}
Now we can express the double logarithmic solution $\Pi(\Gamma_2)$ in terms of the mirror maps $t_i$ for $i=1, \hdots, 4$. For this one has to solve equation \eqref{mirrormaps} for the variables $z_i$ and plug it into $\Pi(\Gamma_2)$. One obtains
\begin{gleichung}
	\Pi(\Gamma_2)		=	\varpi(t_1t_2 + t_1t_3 + t_1t_4 + t_2t_3 + t_2t_4 + t_3t_4)~,
\label{doubleshort}
\end{gleichung}
which is so simple since on a $\mathrm K3$ surface there are no instanton corrections, see also the discussion in section \ref{equalmassK3}.

Again after dividing by the inner point and a transformation into the physical parameters \eqref{batyrevk3} these six basis solutions can be linearly combined to give the maximal cut integral $\mathcal F_{T^3}$ at all points in moduli space.

\subsubsection{Extension to inhomogeneous differential operators}
For the full three-loop banana graph we have to extend the differential operator ideal to an inhomogeneous set of operators. We find these inhomogeneities again when we apply the homogeneous system $\mathcal D_1, \hdots, \mathcal D_4$ on the geometrical differential $\tfrac{u\mu_3}{P_3}$ and perform afterwards an integration over the simplex $\sigma_3$. These integrals can only be performed numerically in all four Batyrev coordinates, but fortunately we can guess their exact values. They are\footnote{Also here we checked this numerically up to more than 15 digits and for different values of the variables $z_i$ for $i=1,2,3$.}
\begin{gleichung}
	\mathcal D_1	\Pi_{\sigma_3}	&=	0 \\
	\mathcal D_2	\Pi_{\sigma_3}	&=	5\log(z_1)-5\log(z_2)	\\
	\mathcal D_3	\Pi_{\sigma_3}	&=	\log(z_1)+ \log(z_2)+ \log(z_3)- 3\log(z_4)	\\
	\mathcal D_4	\Pi_{\sigma_3}	&=	-5\log(z_3)+ 5\log(z_4)~.
\label{inhomk3}
\end{gleichung}

These inhomogeneous differential equations describe all the functions appearing in the Feynman graph \eqref{banana}. The missing special solution can be computed with a triple logarithmic ansatz. For example we can take the following function
\begin{gleichung}
	\varpi_S		=	&-\varpi\left[	\log \left(z_1\right) \log \left(z_2\right) \log \left(z_3\right)+\log \left(z_1\right) \log \left(z_3\right) \log \left(z_4\right)+\log \left(z_1\right) \log \left(z_3\right) \log \left(z_4\right)	\right.	\\
   					&	\left.	+\log \left(z_2\right) \log \left(z_3\right) \log \left(z_4\right)	\right]	 -2\left[	  (z_1+z_2)\left(\log(z_1) + \log(z_2)\right)	+	  (z_1+z_3)\left(\log(z_1) + \log(z_3)\right)	\right. \\
					&	+	  (z_1+z_4)\left(\log(z_1) + \log(z_4)\right)	+	  (z_2+z_3)\left(\log(z_2) + \log(z_3)\right)		\\
					&	\left.+	  (z_2+z_4)\left(\log(z_2) + \log(z_4)\right)	+	  (z_3+z_4)\left(\log(z_2) + \log(z_4)\right)	\right]	\\
					&	+2\left[	(-3z_1+z_2+z_3+z_4)\log(z_1)	+(z_1-3z_2+z_3+z_4)\log(z_2)	\right.	\\
					&	\left.	+(z_1+z_2-3z_3+z_4)\log(z_3)	+(z_1+z_2+z_3-3z_4)\log(z_4)	\right] 	\\
					&	+ 12(z_1+z_2+z_3+z_4)	+ \cdots ~.
\label{specialsolutionk3}
\end{gleichung}

Again, the general solution is then a linear combination of the form $\Pi_{\sigma_3} = \varpi_S + \lambda_0 \varpi + \sum_{i=1}^4\lambda_1^i \Pi(\Gamma_1^i) + \lambda_2\Pi(\Gamma_2)$ with $\lambda_0, \lambda_1^i,\lambda_2 \in \mathbb C$ for $i=1,2,3,4$. We can express $\Pi_{\sigma_3}$ through the physical parameters $(t, \xi_1,\xi_2,\xi_3,\xi_4)$ and divide it by the inner point to yield the full three-loop banana Feynman graph \eqref{banana}.

\subsubsection{The equal mass case and general properties of the ideal of differential operators}
\label{equalmassK3}
For the three-loop banana graph not too many results are known in the literature\footnote{For a discussion on the maximal cut integral in the equal mass case we refere to~\cite{Primo:2017ipr}.}. In the equal mass case there is an inhomogeneous differential equation
{\small{
\begin{gleichung}
	t^2(t-4)(t-16)f_3'''(t) + (6t^3-90t^2+192t)f''_3(t) + (7t^2-68t+64)f'_3(t)	+	(t-4)f_3(t)	=	-4!
\label{opequalbanana}
\end{gleichung}
}}
computed in \cite{mixedhodgestructure}. Restricting our solutions \eqref{regsolutionk3}, \eqref{logsolutions}, \eqref{doublelogsolution} 
and \eqref{specialsolutionk3} to the equal mass case, dividing by the inner point and transform them to the point at infinity in moduli 
space they satisfy equation \eqref{opequalbanana} showing consistency in this limit.

Let us make some general remarks on the properties of the homogeneous  part of the differential operators for periods on K3. 
We first  highlight the structure, which is related to the vanishing string world sheet instantons or unreduced Gromov-Witten 
invariants on K3 manifolds~\cite{MR2746343,MR3524171}, which is expected to hold more generally for hyperk\"ahler 
manifolds. This together with (\ref{specialI})  for $n=2$ and  $r=0,1$ implies a structure for the solutions which is reflected also in the 
classical $W$ invariants of the homogeneous operator ${\cal D}_{K3}$ in  ${\cal D}_{K3} f(t)=-4!$ of (\ref{opequalbanana})  that  determines 
the Feynman graph. To explore the consequences of the vanishing instantons we have to transform the operator for the 
periods $\int_{\Gamma} \Omega$ with $\Omega$ as in (\ref{Omega})  to the point of maximal unipotent monodromy, where 
the instantons are calculated by mirror symmetry in the B-model.    That amounts  to change the variable from $t$ to $z=-1/u$ by  
(\ref{coefficientinnerpoint})  and change the dependent function to $f(z)=f_3 (z)/z$ which yields the  operator 
\begin{equation} 
[\theta^3+ 2 z \theta(1+3\theta +2 \theta^2)-16 z^2( 6+ \theta( 16 + 15 \theta +5 \theta^2) + 96 z^3 (6 +\theta(13+ 9 \theta + 2 \theta^2))]f(z)=0\ .      
\label{k3thetaform}
\end{equation} 
At $z=0$ the unique holomorphic solution is $\varpi=\Pi(T^2)=1+12 z^2-48 z^3 + {\cal O}(z^3)$, while the 
single logarithmic  solution starts with $\Pi(\Gamma_1)=\frac{1}{2 \pi i}[\varpi \log(z)-2 z+ 17 z^2+  {\cal O}(z^3)]$. 
The mirror map is defined as $\tau(z)=\Pi(\Gamma_1)/\Pi(T^2)$ and with $q= \exp(2 \pi i \tau)$ one realises that its inverse is
\begin{equation} 
\frac{1}{z(q)}=\frac{1}{q}- 2 + 15 q -32 q^2 +87 q^3- 192 q^4 + 343 q^5- 672 q^6 +1290 q^7 + {\cal O}(q^8)\ .
\label{mirrormapK3} 
\end{equation} 
This was identified\footnote{Today such identifications of the group and the  $\eta$ quotient for a wide class of groups are 
given by the Webpage of the ``On-line Encylopedia of Integer Sequences''  at {\bf  www//oeis.org} given enough coefficients of series as in (\ref{mirrormapK3}).}  
as $1/z(q)= \left(\frac{\eta(\tau) \eta(3 \tau)}{\eta(2 \tau) \eta(6\tau)}\right)^6+4$ the total modular invariant 
or Hauptmodul of the  group $\Gamma_0(6)^+3$~\cite{verrill1996}. Such identifications have been made for 
many  one-parameter K3 families~\cite{Lian:1995js} based on tables for invariants of Hauptmodules for modular groups  
that features in the monstrous moonshine conjecture~\cite{MR554399}.

Let $\Pi(\Gamma_2)$ be the double logarithmic solution. Because mirror symmetry maps the period vector 
$\Pi^T=(\Pi(T^2),\Pi(\Gamma_1),\Pi(\Gamma_2))$ to the central charges of branes in  integer vertical 
classes $(H_{00},H^{vert}_{11},  H_{22})$ of the mirror  $\mathrm{K3}$, we  can calculate $\Sigma^{ij}$ on the mirror and infer that  
 the $n=2$ and $r=0$ relation in (\ref{specialI})  reads $2 \Pi(T^2) \Pi(\Gamma_2)+ m \Pi(\Gamma_1)^2=0$, where $m$ 
is the self intersection  of the primitive holomorphic curve spanning  $H^{vert}_{11}(M,\mathbb{Z})$. One finds that the period vector 
can be written  as $\Pi^T=\Pi(T^2) (1,\tau,-\frac{m}{2} \tau^2)$. There is also a modular parametrization  of $\Pi(T^2)$ namely 
$z \Pi(T^2)=\frac{(\eta(2\tau) \eta(6 \tau))^4}{(\eta( \tau) \eta(3\tau))^2}$ is the square of periods of a family of elliptic  
curves associated to  $\Gamma_1(6)$. The term  $\frac{m}{2}$  encodes the classical intersection of the mirror $\mathrm{K3}$ and the 
absence of  $q^n$ terms indicates the vanishing of all instanton corrections.     

The classical theory see e.g.~\cite{MR0123757} that goes back to Hermann Schwarz, that  was applied already 
to the one-parameter K3  in~\cite{Lerche:1991wm}, relates the latter fact  to the vanishing of the $W_3$ invariant of the 
$\mathrm{K3}$ operator written generically as  
\begin{equation}   
D f =f'''+3 p(v) f''+ 3 q(v) f'+ r(v) f=0\ .
\end{equation} 
By a change of the dependent function $g(v)=f(v) \exp(\int p dv)$ one eliminates the second derivative            
\begin{equation}
g'''+ 3 Q(v) g'+ R(v) g=0 \  
\end{equation} 
with  $R=r-3 pq + 2p^3-p''$ and  $Q=q-p^2-p'$. Here $Q$  is an invariant of the differential equation, which can be 
used to introduce a new  variable $\tau$, determined as a solution of the Schwarzian equation
\begin{equation} 
\{\tau,v\}=\frac{3}{2} Q \ .
\label{Schwarz} 
\end{equation}  
If the second invariant $W_3=R-\frac{3}{2} Q'=0$ vanishes,  
the function $h=\frac{d \tau }{d v} g$ satisfies the differential equation~\footnote{To prove this one uses  the property $\{x,y\}=-\left( \frac{d x}{d y} \right)^2 \{y,x\}$.} 
 \begin{equation} 
\frac{d^3}{d^3 \tau} h(\tau)=0 \ 
\end{equation}  
with the solution space  $\mathbb{C}\oplus \tau \mathbb{C}\oplus  \tau^2 \mathbb{C}$.   
Schwarz theory determines also the second  order  linear  differential equation 
\begin{equation} 
\frak{D} \frak{f}=\frak{f}''+ 2 \frak{q}(v) \frak{f}' + \frak{q} \frak{f}(v)=0 \ ,
\label{second} 
\end{equation}     
whose ratio of solutions $\tau=\frak{f}_1/\frak{f}_2$ fulfills  (\ref{Schwarz}) and which has the property ${D}={\rm Sym}_2(\frak{D})$, which 
means that  the solutions to $D f=0$ are  $\frak{f}_1^2, \frak{f}_1 \frak{f}_2, \frak{f}_2^2$. It can be found by inverting the following steps:
After the trivial observation that $\frak{g}=\frak{f} e^{\int \frak{p} dv}$ fulfills $\frak{g}''+\frak{Q} \frak{g}=0$, where $\frak{Q}=\frak{q} -\frak{p}^2-\frak{p}'$, 
Schwarz noted that with $\{\tau,v\}=2 \frak{Q}$ defining $\frak{h}= \sqrt{ \frac{d \tau }{d v}}\frak{g}$ the function $\frak{h}$ 
fulfills $\frac{d^2}{d \tau^2} \frak{h}(\tau)=0$ and hence has  solution space  $\mathbb{C}\oplus \tau \mathbb{C}$.

If $\frak{Q}=\frac{3}{4} Q$ then the two  $\tau(v)$ above are identified. Obviously, the solutions $h$ and $g$  are a symmetric 
square of the solutions $\frak{h}$ and $\frak{g}$ respectively and one can arrange $\frak{p}$ so that also the solutions 
$f$ are a symmetric square of the ones of $\frak{f}$. Verrill~\cite{verrill1996} gives this second order equation for  
(\ref{opequalbanana})\footnote{Here $\lambda$ is related to $t$ in (\ref{opequalbanana}) by $\lambda=t-4$.} and~\cite{Broedel:2019kmn} 
relates this  by changes of the dependent and the  independent variable  to the differential  equation for the 
equal  mass sunset graph~(\ref{opequalsunset}).

Four our solutions of the three-loop banana graph with general masses the analogous  structures are 
the equations (\ref{quad}). The first equation together with the vanishing of the genus one worldsheet instantons on 
K3~\cite{MR2746343,MR3524171}, implies the simple form in  (\ref{doubleshort}). The coefficients of the 
double logarithmic terms are  fixed by the intersection theory of the dual curve classes on the mirror $\mathrm{K3}$.
The second equation (\ref{quad}) becomes more powerful in the multi moduli case and restricts the 
structure of the solutions as well as  the differential ideal in  \eqref{l3D1} -- \eqref{l3D4}. One of the 
strongest hints that automorphic forms also gover the maximal cut graph as solution to  
(\ref{l3D1}) -- (\ref{l3D4}) is the mirror map. The analog of (\ref{mirrormapK3})  given as the multi 
parameter inversion of (\ref{logsolutions})  leads  to $1/z_i(q_1,\ldots,q_4)$ for $i=1,\ldots,4$, 
which have also integer expansions in the $q_i=\exp(2\pi it_i)$, where  $t_i=\Pi(\Gamma_1^i)/(2 \pi i \varpi)$
are the K\"ahler parameters of the mirror K3. The natural candidate for these automorphic forms are Borcherds 
lifts of the type discussed in~\cite{MR1625724} and applied to lattice polarized K3 as in~\cite{Klemm:2005pd,Grimm:2007tm}.  
As can be seen from the last two papers  the  automorphic forms are written naturally in terms of the K\"ahler parameters $t_i$ of the mirror.  
The relations to the physical parameters are given by the mirror map defined by (\ref{logsolutions}) and  by (\ref{batyrevk3}).

Finally, let us comment on the higher  loop Banana graphs. For example the analog of the 
differential operator (\ref{k3thetaform}) at suitable large volume coordinates  derives 
analogously from  the $n=5$ entry of Table 1 in  \cite{mixedhodgestructure} as 
(\ref{k3thetaform}) from (\ref{opequalbanana}).  It also appears in the Web database  
explained in \cite{Almkvist,MR3822913} as AESZ34 and is given by
\begin{gleichung} 
	{\cal D}	&= \theta^4-z (35\theta^4+70 \theta^3 +63 \theta^2 + 28 \theta + 5)+ z^2 (\theta+1)^2(259\theta^2 + 518 \theta +285)	\\
				& \quad	-  225 z^3(\theta+1)^2(\theta+2)^2  ~.
\label{DiffBNQ}
\end{gleichung}  
One advantage of the solutions at the MUM point is that  because of the log structure, in case  
a factorization of the solutions exist,  the analytic  solution $\varpi$ must be a pure power of solutions 
of the lower system~\footnote{The easiest way to find  the operator (\ref{second}) on a computer 
might be indeed to take the square root of the unique holomorphic solution $\varpi$ and search for 
a second order operator that annihilates it.}.  If one tries to factorize in this way it will not work. The 
reason can be again understood from (\ref{specialI}), see \cite{MR3965409} for a review. Special  geometry
implies  that the solutions will be $\Pi^T=\Pi(T^3) (1,\tau, \frac{10}{2} \tau^2+{\cal O}(q), -\frac{10}{6} \tau^3+{\cal  O}(q))$
and that $\Pi_3=- \partial_t \Pi_4$.  The reason that this cannot be a symmetric cube are the genus  zero
world sheet instantons encoded in the higher series in $q$. For this geometry of the one-parameter family of  
Barth-Nieto quintics they are not vanishing to all degrees. Subtracting  the multi-covering  
contributions the first $n_d^{(0)}\in \mathbb{Z}$  are given for degree $d=1,\ldots,7$ by $24,48,224,1248,8400,62816,516336$. 
Despite the integer structures in the $n_d^{(0)}$ and the mirror map $1/z=1/q + 8 + 28 q + 104 q^2 + 654 q^3 +{\cal O}(q^4)$
it will be much more complicated to give closed  automorphic expressions for the equal mass 
four-loop graph then for the general mass three-loop graph.

There are however interesting relations of the periods  to modular forms of $\Gamma_0(N)$ and algebraic extensions at the rank  two  attractor points  that (\ref{DiffBNQ}) 
as studied in \cite{CDEV}. At these points the numerator of the  Hasse Weil  factorises and  the exact values of maximal cut integral  
are given by $L$-function values of holomorphic Hecke Eigenforms  forms of weight two and four of $\Gamma_0(N)$~\cite{CDEV} 
or extensions  and the quasi-periods of the corresponding meromorphic forms~\cite{AED}\cite{BK}.

\section{Conclusions and Outlook}
\label{conclusions}

The geometric interpretation relating  Feynman integrals to Calabi-Yau chain integrals leads to powerful new 
calculational methods. In particular, the resonant GKZ differential  system that was used in the context of 
mirror symmetry to the period integrals of  Calabi-Yau hypersurfaces in toric varities~\cite{MR1269718,Hosono:1993qy,Hosono:1995bm}  
yields  straightforwardly to  the maximal cut integral at the point of maximal unipotent  monodromy.  The advantage 
of the GKZ differential system is that it uses the symmetries of the Newton polytopes associated to the  banana graphs most efficiently. Its  disadvantage, namely that it has more 
solutions and more variables than the actual  Calabi-Yau and Feynman integrals, can be overcome using methods from the mirror symmetry application of the 
GKZ system~\cite{MR1269718,Hosono:1993qy,Hosono:1995bm}. The latter allows us to derive the complete homogeneous Picard-Fuchs differential  ideal in 
the physical parameters. The solutions to this  differential ideal  characterizes the analytic form of the maximal cut integral  
everywhere in the physical parameter space. The use of the symmetries in this approach turns out to be more efficient 
than the multi parameter Griffiths reduction method. Such relations between master integrals for different classes of Feynman graphs appear in the physics literature in\cite{Frellesvig:2017aai,Bosma:2017ens,Harley:2017qut}.

Moreover, at the point of maximal unipotent monodromy we could  determine the inhomogeneity 
by integrating directly the geometrical chain integral  after applying the generators 
of the  homogeneous Picard-Fuchs differential  ideal  to its integrand. The form of the  corresponding 
inhomogeneities turn out to be very simply. This allows us to find an inhomogeneous solution and 
express for  the first time the full mass dependence of the three-loop banana graph analytically. The 
result is related to the chain integrals that appear in the calculation  of open  topological string 
amplitudes.

The GKZ integrals and Feynman integrals can have more general rational functions as integrand  
than the simple one that is realized for the Banana graph. The scaling invariance  that occur  in 
Feynman integrals, lead however typically to  GKZ systems related to Calabi-Yau geometries. However,  
their desingularizations can have much more complicated realizations as the hypersurfaces in toric varieties 
that feature in this paper. For instance, complete intersections in toric varities or even more exotic cases as  
Paffian Calabi-Yau spaces in Grassmanians or  flag manifolds are conceivable at least in special slices
of the moduli space. Nevertheless, we expect that many aspects of the general approach outlined 
in this paper should apply. In particular, the GKZ system has been applied to the complete 
intersection three-fold case in~\cite{Hosono:1994ax,MR1328251,MR1463173} and to higher dimensional  Calabi-Yau 
manifolds  in~\cite{Greene:1993vm,Mayr:1996sh,Klemm:1996ts,Bizet:2014uua}.
Recently, progress has been made concerning the more exotic realizations of Calabi-Yau spaces in the $(2,2)$ 
supersymmetric  2d gauge linear $\sigma$ model approach with non-abelian gauge groups. For example  
in~\cite{Gerhardus:2016iot,Gerhardus:2018zwb} the  Picard-Fuchs  operators for such geometries 
have been obtained using  localization techniques.

Moreover, there are important universal properties that govern the Calabi-Yau periods   
completely independent of their geometrical realization. In particular, there are the 
{\em transversality identities} (\ref{transversal}) which have fundamental consequences 
on the period geometry of Calabi-Yau manifolds, which are very different in even and odd 
dimensions. Together with some likewise universal properties about the integrality
of the mirror map as well as the integrality of instantons and vanishing theorems  for the 
latter, it strongly restricts the classes of automorphic functions that can encode the 
Feynman integrals.

Our main result is the calculation of the three-loop graph. Let us shortly comment on the possibility to 
extend our methods to the four loop banana graph: It  is possible to find the analogs of \eqref{batybubble}, \eqref{batyrevsunset} and \eqref{batyrevk3} 
as well  as of the differential ideals  (\ref{opbubble}), (\ref{opsunset}),  and (\ref{l3D1}) -- (\ref{l3D4}) 
for the four-loop graph. Also the inhomogeneous terms \eqref{inhombubble}, \eqref{inhomsunset} and 
(\ref{inhomk3}) are expected to generalize. With some efforts to code the recursions  that 
follow form the analog  of   (\ref{l3D1}) -- (\ref{l3D4}) as well as (\ref{specialsolutionk3}) efficiently, 
it should be possible  to find fast convergent expressions  for the  four-loop  general 
mass case, just as it is possible for the three-loop case.

\vspace{1.5cm}
\section*{Acknowledgements}
We like to thank Mahsa Barzegar, Kilian B\"onisch, Ruth Britto, Francis Brown,  Philip Candelas, Xenia del la Ossa, Claude Duhr, Mohamed Elmi,  Hans Jockers, Rene Klausen,  Duco van Straten, 
Emanuel Scheidegger and  Don Zagier for discussion on various aspects of this work  as well as very enlightening talks related to the subject.  AK wants to thank the LPT-ENS 
for hospitality and thank Pierre Vanhove for discussions, which have triggered this project.

\appendix

\section{Differential Operator Ideal of the Banana Graph}
\label{appop}

Here we list a generating set of differential operators which describes the three-loop banana graph in all four physically important Batyrev coordinates.

\vspace{0.25cm}
\begin{fleqn}
\begin{gleichung}
	\mathcal D_1	&=	\left(\theta _1-\theta _2\right) \left(\theta _3-\theta _4\right)		\\
				&\quad +z_1(\theta_3-\theta_4)(\theta_1-\theta_2-\theta_3-\theta_4)		+	z_2(\theta_3-\theta_4)(\theta_1-\theta_2+\theta_3+\theta_4)	\\
				&\quad -2(z_1-z_2)\left(	z_3(\theta_3+1)-z_4(\theta_4+1)	\right)(\theta_1+\theta_2+\theta_3+\theta_4+1)
\label{l3D1} 
\end{gleichung}
\end{fleqn}

\begin{fleqn}
\begin{gleichung}
	\mathcal D_2	&=	5(\theta_1-\theta_2)\theta_4-6\theta_2^2	\\
				&\quad +z_1 \left(	2 \theta _1^2-8 \theta _1 \theta _2+6 \theta _2^2-6 \theta _3^2-11 \theta _4^2+4
   \left(\theta _1+\theta _2\right) \theta _3+\left(9 \theta _1-\theta _2-13 \theta_3\right) \theta _4	\right)	\\
   				&\quad +z_2 \left(	17 \theta _4^2+\left(13 \theta _1-9 \theta _2+25 \theta _3+6\right) \theta _4-2
   \left(\theta _2-\theta _3\right) \left(4 \theta _2+6 \theta _3+3\right)+\theta _1   \left(8 \theta _2+8 \theta _3+6\right)	\right)	\\
   				&\quad +	2\left[	5z_3z_4(\theta_2-\theta_1)+z_1^2(\theta_1-\theta_2-\theta_3-\theta_4)+z_2^2(\theta_1-\theta_2+\theta_3+\theta_4) \right.	\\
				&\quad \left.	+z_1z_4(3 \theta _1+3 \theta _2-2 \theta _3-8 \theta _4-5)	+z_1z_3(3 \left(\theta _1+\theta _2-\theta _3\right)-2 \theta _4)	\right.	\\
				&\quad \left.	+3z_1z_2(-\theta _1+3 \theta _2+\theta _3+\theta _4+2)+z_2z_3(6 \theta _3+5 \theta _4+6)	\right.	\\
				&\quad \left.	+z_2z_4(5 \theta _3+11 \theta _4+11)	\right] (\theta _1+\theta _2+\theta _3+\theta _4+1)
\label{l3D2} 				
\end{gleichung}
\end{fleqn}
\vspace{-0.75cm}
\begin{fleqn}
\begin{gleichung}
	\mathcal D_3	&=	-3 \theta _2^2-2 \theta _2 \theta _4+\theta _1 \left(3 \theta _2-2 \theta
   _4\right)+\theta _4 \left(\theta _3+\theta _4\right)	\\
   				&\quad	-3z_1\theta _2 \left(-\theta _1+\theta _2+\theta _3\right)-z_1\theta_4(2 \theta _1+\theta _2-2 \theta _3)-z_3\theta_4(\theta _1+\theta _2-\theta _3)+(2z_1-z_3)\theta_4^2		\\
				&\quad -z_4\left(\theta _1+\theta _2+\theta _3-\theta _4\right) \left(\theta _4+1\right) +z_2\left(\theta _1-\theta _2+\theta _3+\theta _4\right) \left(3 \theta _2+8 \theta_4+3\right)	\\
				&\quad +2\left[	-2z_3z_4(\theta_4+1)+z_1z_4-3z_1z_3\theta_2+z_1(z_3+z_4)\theta_4 +z_2z_3(3 \theta _2+4 \theta _4+3)		\right.	\\
				&\quad \left.	+4z_2z_4	+4z_2(z_1+z_4)\theta_4	\right]	(\theta _1+\theta _2+\theta _3+\theta _4+1)
\label{l3D3} 				
\end{gleichung}
\end{fleqn}
\vspace{-0.75cm}
\begin{fleqn}
\begin{gleichung}
	\mathcal D_4	&=	-\theta _2 \left(\theta _2+5 \theta _3-5 \theta _4\right)		\\
				&\quad	+z_1(2 \theta _1^2-\left(3 \theta _2+\theta _3-4 \theta _4\right) \theta _1+\theta _2^2-\theta
   _3^2-6 \theta _4^2+4 \theta _2 \theta _3-\left(\theta _2+3 \theta _3\right) \theta _4)		\\
   				&\quad	+5z_4\left(\theta _1-\theta _2-\theta _3\right) \left(\theta _1+\theta _2+\theta _3-\theta
   _4\right) +5z_3\theta _4 \left(\theta _1+\theta _2-\theta _3+\theta _4\right)	\\
   				&\quad	+z_2 \left[	-3 \theta _2^2+\left(-14 \theta _3+11 \theta _4-1\right) \theta _2+17 \theta _3^2-8
   \theta _4^2+\theta _3+\theta _1 \left(3 \theta _2+13 \theta _3-12 \theta _4+1\right)	\right.	\\
   				&\quad	\left.	+5\theta _3 \theta _4+\theta _4	\right]	\\
				&\quad	+\left[	2z_1^2(\theta _1-\theta _2-\theta _3-\theta _4)	+z_1z_4(11 \theta _1-9 \theta _2+\theta _3-11 \theta _4)	\right.	\\
				&\quad	\left.	+z_1z_2(-\theta _1+3 \theta _2+11 \theta _3-9 \theta _4+2)+z_1z_3(\theta _1+11 \theta _2-\theta _3+\theta _4) \right.	\\
				&\quad	\left.	+2z_2z_3(-5 \theta _2+11 \theta _3-5 \theta _4+6)+2z_2z_4(5 \theta _3-4 \theta _4-4)+10z_3z_4(\theta _4-\theta _3)	\right.	\\
				&\quad	\left.	+2z_2^2	(\theta _1-\theta _2+\theta _3+\theta _4)	\right]	(\theta _1+\theta _2+\theta _3+\theta _4+1)
\label{l3D4}				
\end{gleichung}
\end{fleqn}

\newpage

\addcontentsline{toc}{section}{References}
\bibliographystyle{utphys}
\bibliography{pqm}

\end{document}